\documentclass[12pt]{article}
\usepackage[tbtags]{amsmath}
\usepackage{amssymb}
\usepackage{amsthm}
\usepackage{array}
\usepackage{xy}
\usepackage{slashed}

\usepackage[iso-8859-7]{inputenc}
\usepackage{graphicx}

\def\<{\left\langle}
\def\>{\right\rangle}

\title{\bfseries Suppressing gravitino thermal production with a temperature-dependent messenger coupling \normalfont}
\author{Marcin Badziak,$^1$ 
\,\, Ioannis Dalianis,$^2$ 
 and \,Zygmunt Lalak$^1$ 
\\
$^1$ \it Institute of Theoretical Physics, Faculty of Physics, University of Warsaw\\
\it ul.\ Pasteura 5, PL--02--093 Warsaw, Poland\\
$^2$ \it Physics Division, National Technical University of Athens, \\ \it 15780 Zografou Campus, Athens, Greece}

\begin{document}


\newcommand{\g}{\greektext} 
\newcommand{\e}{\latintext}
\maketitle
\abstract{ 
We show that the constraints on GMSB theories from the gravitino cosmology can be significantly relaxed if the messenger-spurion coupling is temperature dependent. We demonstrate this novel mechanism in a scenario in which this coupling depends on 
the VEV of an extra singlet field $S$ that interacts with the thermalized plasma which can result in a significantly suppressed  gravitino production rate. 
In such a scenario the relic gravitino abundance is determined by the thermal dynamics of the $S$ field and it is easy 
to fit the observed dark matter abundance evading the stringent constraints on the reheating temperature, thus making gravitino dark matter consistent with
thermal leptogenesis.}

\begin{sloppypar}
	
\end{sloppypar}

\tableofcontents

\section{Introduction}

\subsection{Overview of the GMSB gravitino cosmology}
The known particles in nature cannot account for the dark matter in the galaxies. Physics beyond the Standard Model is required. Supersymmetry is one of the most motivated theories that is expected to operate at the TeV scale. A great advantage for supersymmetry is that it provides excellent dark matter candidates \cite{Pagels:1981ke}. 
In the Gauge Mediated Supersymmetry Breaking scenario (GMSB) \cite{Dine:1981za, Dimopoulos:1981au, Dine:1981gu, AlvarezGaume:1981wy, Dine:1982zb, Dimopoulos:1982gm,Nappi:1982hm, Dine:1993yw}
the Lightest Supersymmetric Particle (LSP), is the gravitino which, if stable, is an ultra-weakly interacting particle. 

If the gravitino is part of the dark matter in the universe it must have a relic abundance $\Omega_{3/2}h^2 \lesssim 0.12$ \cite{Ade:2015xua}. It is supposed to be generated at the early hot universe from scatterings and decays in the thermal plasma as well as from non-thermal decays. 
In a minimal set up where the thermal plasma consists of the Standard Model (SM) particles plus their superpartners (MSSM)  \cite{Drees:2004jm} the gravitino is mainly generated from thermal scatterings of the MSSM degrees of freedom. 
In such a case
the relic abundance of the helicity $\pm1/2$ states is given by \cite{Bolz:2000fu, Pradler:2006qh, Rychkov:2007uq}
\begin{equation} \label{MSSM}
\Omega^\text{MSSM(sc)}_{3/2}h^2 \sim 0.1 \left( \frac{T_\text{rh}}{10^8 \,\text{GeV}}\right) \left(\frac{\text{GeV}}{m_{3/2}} \right) \left(\frac{m_{\tilde{g}}}{\text{TeV}}\right)^2\,,
\end{equation}
where $m_{\tilde{g}}$, $m_{3/2}$ the gluino and gravitino masses respectively and $T_\text{rh}$ the reheating temperature.

According to the expression (\ref{MSSM}), the constraint $\Omega_{3/2}h^2 \leq 0.12$ implies a bound on the maximum reheating temperatures for a particular gravitino mass \cite{Moroi:1993mb}. A stringent bound on the reheating temperature can prevent the generation of the baryon asymmetry of the universe through leptogenesis \cite{Fukugita:1986hr, Buchmuller:1999cu, Davidson:2002qv, Davidson:2008bu}. Gravitinos with light masses, $m_{3/2}\lesssim 1 $ keV can attain a thermal equilibrium abundance without saturating the relic density constraint and thus escape the maximimum reheating temperature bound. However,  unless the gravitino is ultralight, $m_{3/2}< 16$ eV, a light gravitino is a warm dark matter component and excluded by the Lyman-$\alpha$ forest data \cite{Viel:2005qj}.

For a specific gravitino mass, tuning the reheating temperature to a specific value can be an explanation for the gravitino dark matter abundance. However, this is not a generic solution. The LHC lower bounds on the gluino masses have shortened further the parameter space where successful fine tuning can take place, see e.g. ref. \cite{Knapen:2015qba, Roszkowski:2012nq}. In addition, scenarios such as the thermal leptogenesis, that requires high reheating temperatures, is hard to be reconciled with viable gravitino cosmology. R-parity violating vacua \cite{Buchmuller:2007ui} or late entropy production \cite{Fujii:2002fv, Fukushima:2013vxa} has to be invoked to address the tension. The constraints become much more severe when the messengers fields, the building blocks of the GMSB theories, are involved in the gravitino generation processes \cite{Choi:1999xm, Jedamzik:2005ir, Dalianis:2013pya, Fukushima:2012ra,Fukushima:2013vxa}.

The GMSB secluded sector is crucial because the gravitino obtains its mass from the superhiggs mechanism and at energy scales $E\gg m_{3/2}$ the helicity $\pm1/2$ component is identified with the Goldstino field $\psi_G$, the massless fermion associated with the supersymmetry breaking. 
The chiral superfield $X$ parametrizes the supersymmetry breaking 
via its $F$-term non-vanishing value
\begin{equation}
\left\langle X \right\rangle = X_0 + \theta^2\, F \,
\end{equation}
and the $\psi_G$ field is the fermionic component of the $X$ superfield that corresponds to the helicity $\pm1/2$ gravitino component. Also, the value of the scalar component of $X$ breaks the $R$-symmetry and non vanishing masses for gauginos can be generated. Therefore,  estimating the relic gravitino abundance from the MSSM sector 
cannot account for a complete estimation. The GMSB secluded sector dynamics plays a major r\^ole and needs to be 
taken into account as well. 

Messenger fields of the GMSB models transmit the supersymmetry breaking efficiently to the charged Standard Model (SM) fields and naturally the gravitino of the supergravity multiplet is the LSP.
The simplest GMSB set up is described by the following superpotential
\begin{equation} \label{secl}
W_\text{mess}= \lambda_\text{mess} X \phi \bar{\phi}\,,
\end{equation}
where $\phi$, $\bar{\phi}$ the messenger superfields. The messengers can belong in some GUT representation that contains the SM gauge group.  
Assuming that the global R symmetry is broken at the vacuum the gaugino masses are given by the formula 
\begin{equation} \label{soft}
m_{\tilde{g}_i} = \frac{\alpha_i}{4\pi} \frac{\lambda_\text{mess}\, F}{M_\text{mess}}\,,
\end{equation}
where the index $i$ runs over the SM gauge groups and $M_\text{mess}$ is the mass of the messenger fields. 

The fermionic components of the messenger superfields have the mass $M_\text{mess}$, while the scalar messengers have the masses squared $M^2_\text{mess} \pm \lambda_\text{mess} F$.  The collider constraints on the gaugino masses put a lower bound on the ratio $\lambda_\text{mess} \, F/ M_\text{mess} >10^5$ GeV. This implies that the messenger mass parameter is
\begin{equation}\
M_\text{mess} > 10^5\, \text{GeV}
\end{equation}
in order to avoid electroweak and color breaking in the messenger sector.

\begin{figure} 
\centering
\includegraphics [scale=.9, angle=0]{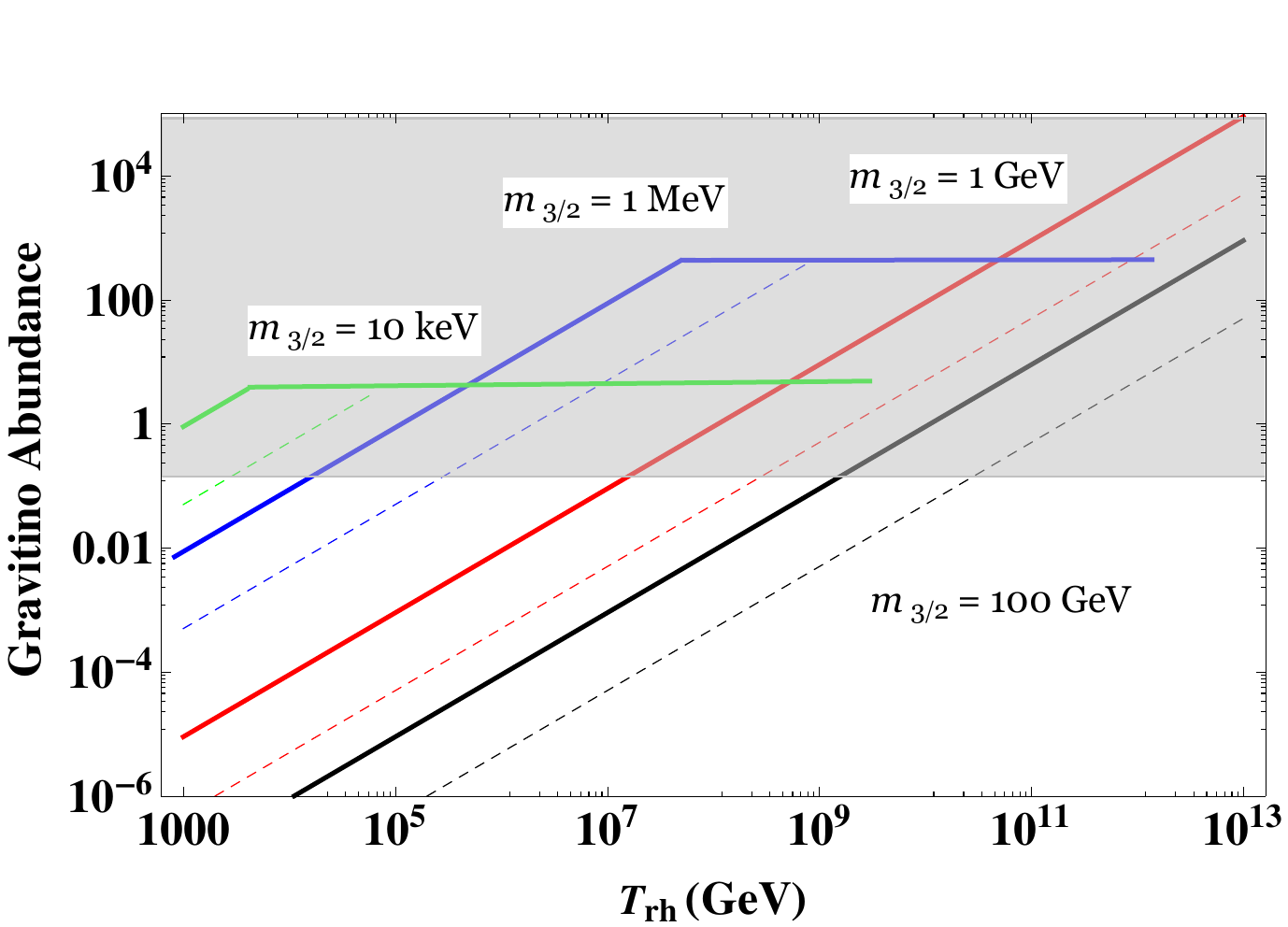} 
\caption {\small {The allowed regions in the contour ($\Omega_{3/2}h^2, T_\text{rh}$), for $m_{\tilde{g}}=700$ GeV (dashed) and $m_{\tilde{g}}=3$ TeV (solid) and $M_\text{mess}>T_\text{rh}$. From bottom to the top, the pair of lines correspond to $m_{3/2}=100$ GeV, 1 GeV, 1 MeV and 10 keV. The plateau in the abundance of the light gravitinos is due to their thermalization. 
}}
\end{figure}

When the GMSB sector is taken into account the gravitino relic abundance result, expressed by the formula (\ref{MSSM}), radically changes. 
The thermalized messengers affect both the vacuum structure of the theory and gravitino production rate. There are two remarks to be underlined:
\begin{itemize}
\item
For $T>M_\text{mess}$ the messengers become relativistic, are integrated in, and it has been shown that the gravitino production rate from the MSSM is suppressed by the factor $M^2_\text{mess}/T^2$ compared to the conventional rate one obtains for energies lower than the messenger mass scale   \cite{Choi:1999xm, Fukushima:2012ra,Fukushima:2013vxa}.  
\item
Thermal effects can restore the global $R$-symmetry at the temperature $T_{\not{R}}$. 
A hierarchy between the sfermion and gaugino masses is obtained during this $R$-symmetric thermal phase.
The $R$-suppressed gaugino masses imply a suppressed gravitino production rate from the MSSM \cite{Dalianis:2011ic}. 
\end{itemize}
Except if $T_{\not{R}} < M_\text{mess}$, the MSSM efficiently generates gravitinos for temperatures below the $M_\text{mess}$ scale, and this is the case we consider in this paper. These observations are not accounted into formula (\ref{MSSM}). 
Instead the formula should be corrected to \cite{Choi:1999xm, Fukushima:2012ra,Fukushima:2013vxa}  
\begin{equation} \label{mssm}
\Omega^\text{MSSM(sc)}_{3/2} \, h^2\sim
\begin{cases} 
 0.1 \left( \frac{M_\text{mess}}{10^8 \,\text{GeV}}\right) \left(\frac{\text{GeV}}{m_{3/2}} \right) \left(\frac{m_{\tilde{g}}}{\text{TeV}}\right)^2 &  T_\text{rh} \gtrsim M_\text{mess} \\ 
0.1 \left( \frac{T_\text{rh}}{10^8 \,\text{GeV}}\right) \left(\frac{\text{GeV}}{m_{3/2}} \right) \left(\frac{m_{\tilde{g}}}{\text{TeV}}\right)^2  & T_\text{rh} \lesssim M_\text{mess} \,
\end{cases}\,\,\,
\end{equation}
We see that the expected relation $\Omega^\text{MSSM(sc)}_{3/2} \propto T_\text{rh}$ is not generic: above the messenger scale the generation of gravitinos from the MSSM sector effectively ceases.

Apart from the fact that the dynamics of the thermalized GMSB sector intervene in the production of gravitinos from the MSSM, additional scatterings of thermalized messengers contribute to the gravitino yield. To this end, an extra term has to be added at (\ref{mssm}) which actually dominates over the one coming from the MSSM contribution for $T_\text{rh}> M_\text{mess}$,
\begin{equation} \label{mess1}
\Omega^\text{mess(sc)}_{3/2} \, h^2\sim  0.4 \, \left(\frac{M_\text{mess}}{10^4 \,\text{GeV}}\right) \left(\frac{\text{GeV}}{m_{3/2}} \right) \left(\frac{m_{\tilde{g}}}{\text{TeV}}\right)^2\,.
\end{equation}
Unless $T_\text{rh}<M_\text{mess}$ or a late entropy production takes place, the formula (\ref{mess1}) manifests that messenger masses $M_\text{mess} > 10^5$GeV rule out the gravitino LSP scenario with $16 \text{eV} \lesssim m_{3/2}\lesssim (10-100)$ GeV.  There is only a small window of low scale messenger mass values and large gravitino masses that do not constrain the reheating temperature \cite{Dalianis:2013pya}, see figure 2.   
Moreover, in ref. \cite{Dalianis:2013pya} it is mentioned  that the Goldstinos acquire a thermal equilibrium distribution
\begin{equation} \label{Oth}
\left. \Omega^{\text{eq}}_{3/2}h^2 \simeq {\cal O} (5)\times 10^{5} \left( \frac{m_{3/2}}{\text{GeV}} \right) \left[\frac{270}{g_*(T)} \right]\right|_{T=M_\text{mess}} 
\end{equation}
for moderate values for the messenger coupling 
\begin{equation} \label{mgeq}
\left. \lambda_\text{mess} \, > \sqrt{ \frac{T}{M_\text{Pl}}} \, \right|_{T=M_\text{mess}}\simeq 10^{-5} \left( \frac{M_\text{mess}}{10^8 \text{GeV}}\right)^{1/2}\,.
\end{equation}
Summing up, without any late time dilution, the gravitino abundance reads  
\begin{equation}
\Omega^\text{tot}_{3/2}  \, =\, \Omega^\text{MSSM(sc)}_{3/2} \, +\, \Omega^\text{mess(sc)}_{3/2}\, + \,\Omega^\text{n-th}_{3/2}  
\end{equation}
where the "n-th" stands for the non-thermal contribution coming from the GMSB and MSSM field decays that we will mostly neglect as subdominant in the subsequent analysis. There are scenarios that the non-thermal contribution is a major one, see e.g ref. \cite{Fukushima:2013vxa} and \cite{Eberl:2015dia} for recent studies on non-thermal production of gravitinos from the GMSB and MSSM sector respectively.

\begin{figure} 
\centering
\begin{tabular}{cc}
{(a)} \includegraphics [scale=.5, angle=0]{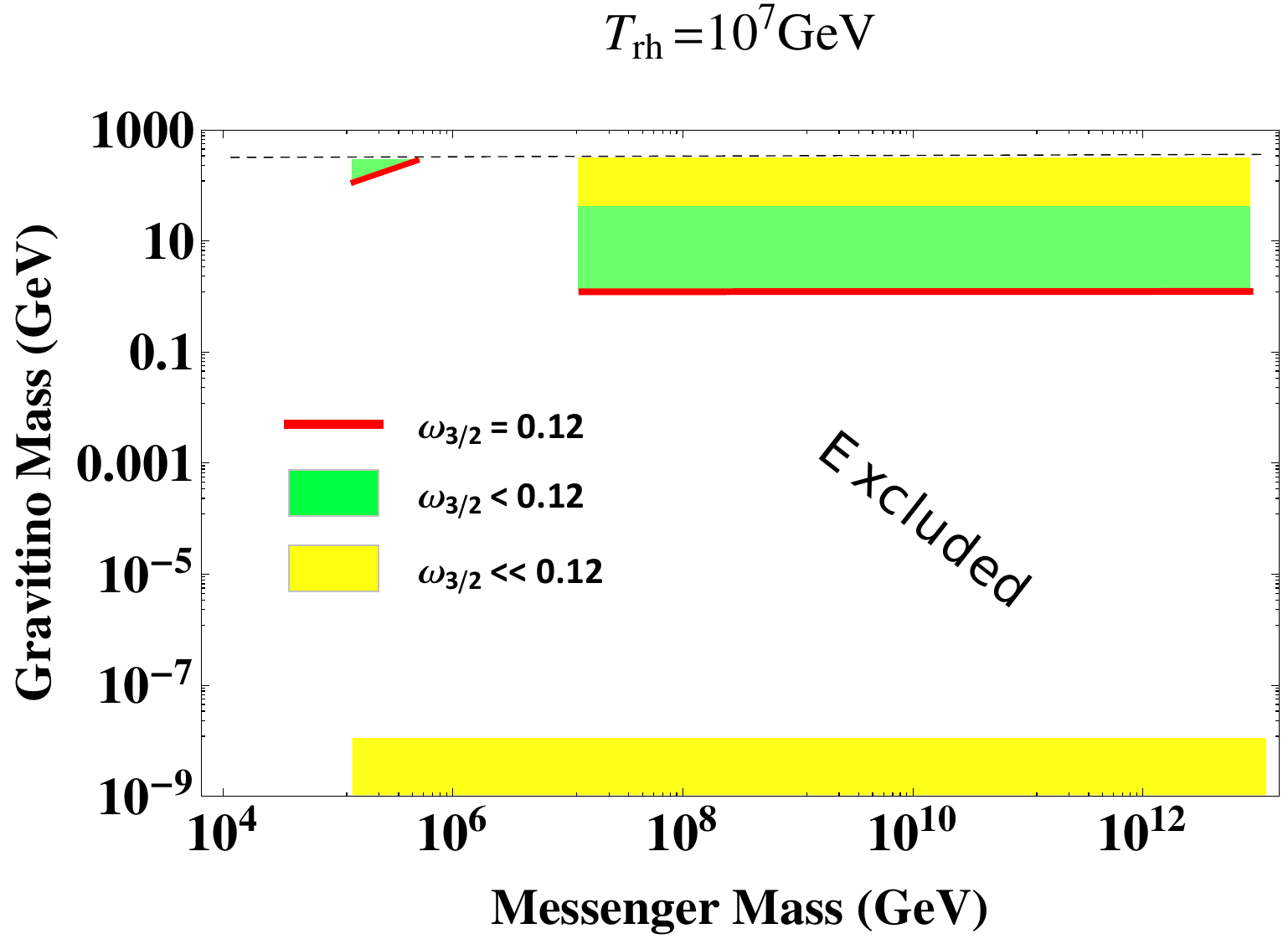} &
{(b)} \includegraphics [scale=.5, angle=0]{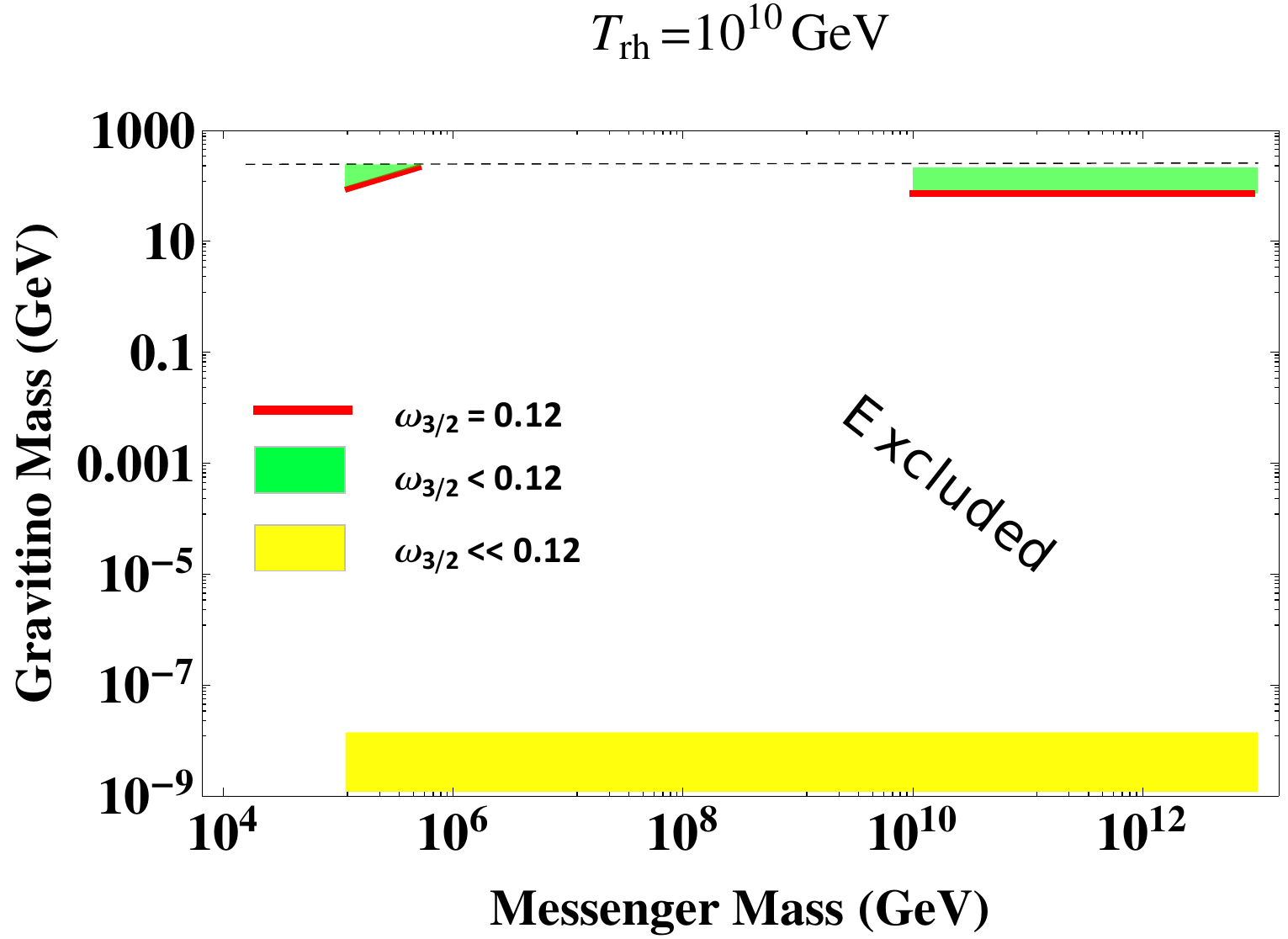}  \\
\end{tabular}
\caption{\small{ The allowed regions of gravitino abundance $\omega_{3/2}\equiv \Omega_{3/2} h^2$ in the contour ($m_{3/2}, M_\text{mess}$), for $m_{\tilde{g}}=3$ TeV, $M_\text{mess}>10^5$ GeV and without any late entropy production.}}
\end{figure}

The above results stress out that the messenger sector is the regulator of the gravitino abundance in the GMSB scenarios and mainly responsible for the gravitino overabundnace problem.
One can choose specific set of values for the key parameters $T_\text{rh}$, $m_{3/2}$ and $M_\text{mess}$ that admit viable gravitino LSP cosmology however, it is of much interest to construct models where this tuning could be avoided.  A mechanism that can reconcile the cosmology of GMSB models with low messenger mass scale, which are particularly attractive phenomenologically due to naturalness arguments -see e.g. ref. \cite{Craig:2012xp}, and thermal leptogenesis is certainly advantageous.

\subsection{The model: field dependent messenger coupling}

In this work we suggest that the gravitino yield in the early universe can change dramatically if the value of the messenger coupling diminishes during the thermal phase due to collective phenomena. At finite temperature the limit $\lambda_\text{mess} \rightarrow 0$ implies that the messengers effectively decouple from the spurion $X$ and the mass splitting inside their multiplets vanishes. Thereby, the Goldstino decouples form the plasma and 
the production rate of Goldstinos gets significantly suppressed.
Evidently, when the temperature drops below a critical value, $T_S$, the messenger Yukawa coupling value gets restored.

Such a scenario can be implemented if the messenger coupling is controlled by the VEV of an extra singlet field $S$ 
\begin{equation}
W_\text{mess}= \lambda_\text{mess}(S)X\phi\bar\phi\,.
\end{equation} 
Here the messenger superpotential is comprised of a higher dimensional operator 
\begin{equation}
\lambda_\text{mess}(S)= \left(\frac{S}{\Lambda_*}\right)^n\,.
\end{equation}
The $\Lambda_*$  is the cut-off scale of the non-renormalizable interaction and we consider $\Lambda_*>M_\text{mess}$. 
The scalar singlet is assumed to be stabilized in a non-zero VEV $S_0$. Due to interactions, with other fields $\varphi, \bar{\varphi}$ that participate in the thermal equilibrium, that may be described by the superpotential  
\begin{equation} \label{11}
\delta W= \kappa S \varphi \bar{\varphi}\,,
\end{equation}
 a thermal mass $\sim\kappa T$ is induced for the singlet. 
These thermal effects are capable to drive the singlet to the origin of the field space where the messenger coupling vanishes and the Goldstino decouples.
When the temperature falls below a critical value $T_S$, characteristic for the transition of the singlet towards the zero temperature vacuum $S_0$, the efficiency of the gravitino production rate is recovered since  $\lambda_\text{mess}(T=T_S)\simeq \lambda_{\text{mess}}(T=0)$. If the $T_S$ is below the messenger mass scale the contribution from the thermal scattering of messengers is vanishing and the Goldstino production takes place only via the thermal scatterings in the MSSM plasma.

The alteration of the gravitino production rate is the outcome of the dynamics of $S$ at finite temperature. 
The $\lambda_\text{mess}(T)\rightarrow 0$ is realized only if the singlet remains at the vacuum state and does not get thermally excited.  This fact constrains the couplings of the singlet to the thermalized degrees of freedom to be very weak $\kappa\ll 1$. Otherwise, if the singlet is tightly coupled to the plasma and fast thermalized, the gravitino production rate can be actually amplified. Nonetheless, a very weak coupling to the plasma is welcome because it prevents a long-lasting trapping of the singlet at the origin and thus an $S$-domination phase from taking place.

An interesting and workable scenario is to identify the Higgs bilinear in the place of the unspecified $\varphi, \bar{\varphi}$ fields in the superpotential (\ref{11}). Then the $S$ plays the r\^ole either of the Peccei-Quinn saxion or of the NMSSM singlet. In this context the singlet VEV $S_0$ can determine both the value of the messenger coupling and the $\mu/B_{\mu}$ ratio.
\\
\\
In the next section we derive the Goldstino yield at finite temperature when the messenger coupling is $\lambda_\text{mess}(S)=(S/\Lambda_*)^n$. 
In section 3 we discuss the superpotential coupling of the singlet to the Higgs bilinear and finally in section 4 we conclude. Some technical details concerning the computation of the Goldstino yield and elements of the gravitino cosmology can be found in the appendix. Throughout the text we will often use the Goldstino $\psi_G$ to refer to the gravitino $\tilde{G}$ since at energy scales much larger than $m_{3/2}$ the equivalence theorem applies. Also, we will use the same symbol $S$ for the singlet superfield and its scalar component.


\section{Thermal production of Goldstinos for $\lambda_\text{mess}(T)$}

\begin{figure} 
\centering
\begin{tabular}{cc}
{(a)} \includegraphics [scale=.5, angle=0]{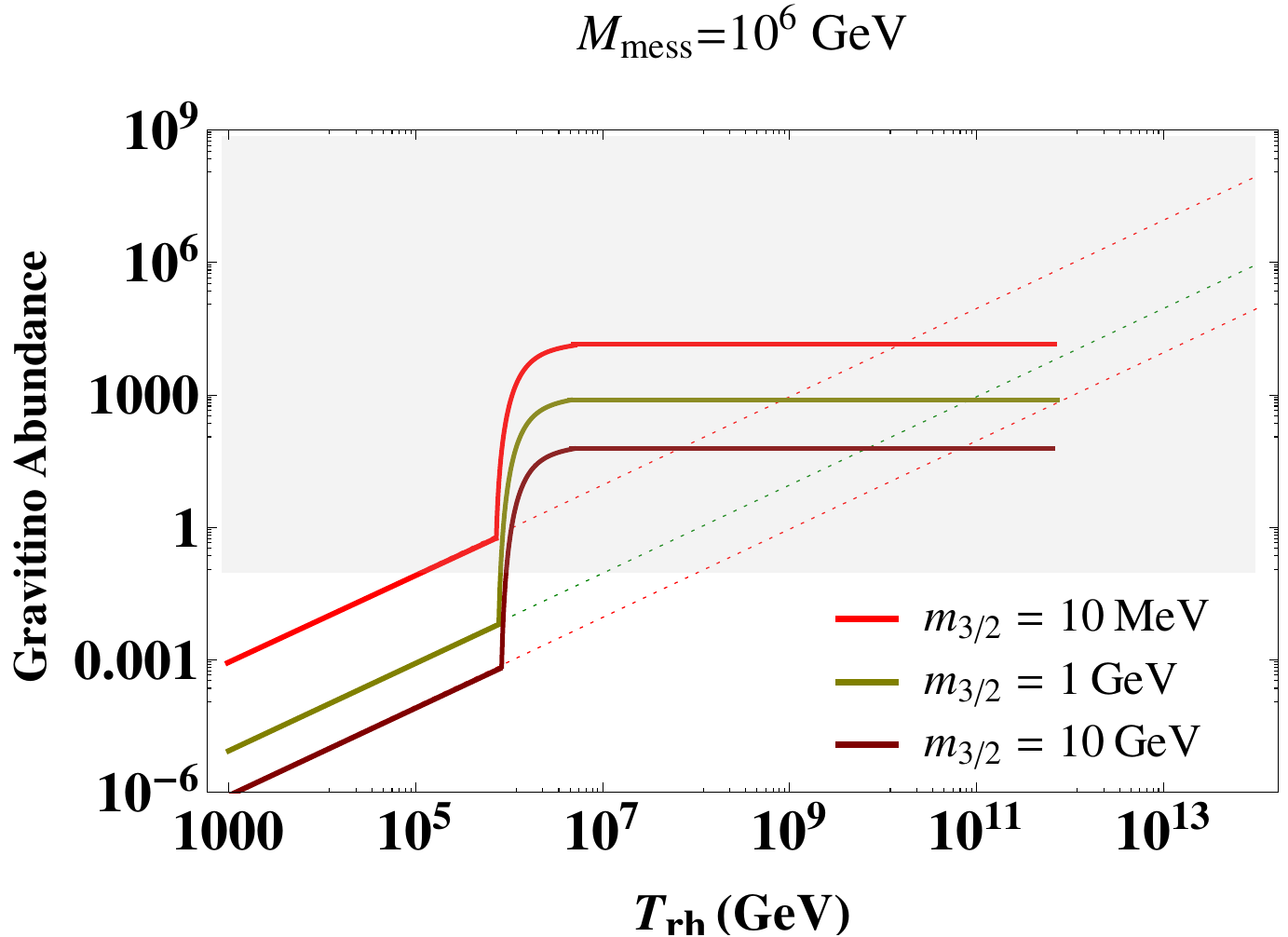} &
{(b)} \includegraphics [scale=.5, angle=0]{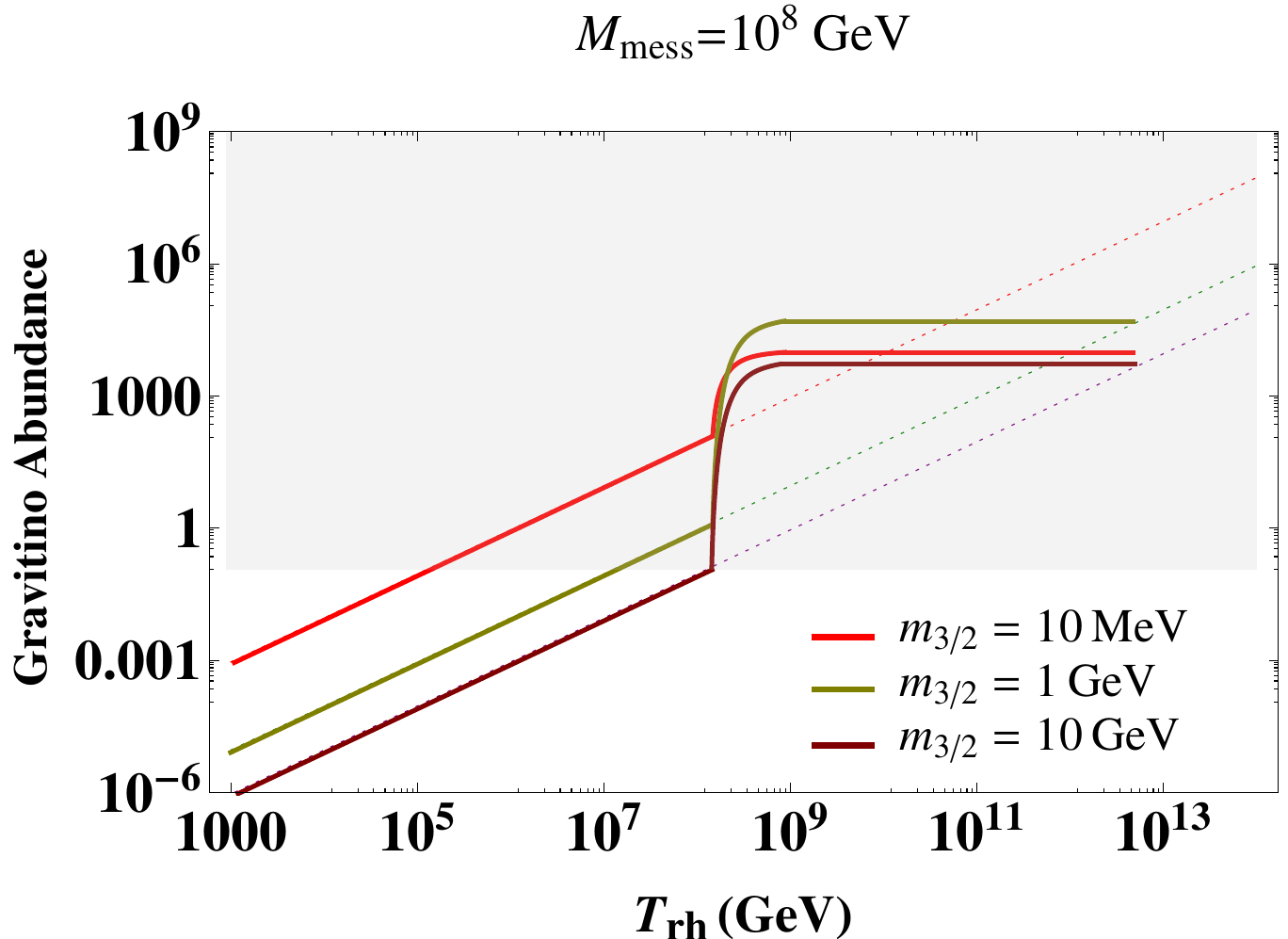}  \\
\end{tabular}
\caption{\small{ The dependence of the gravitino abundance $\Omega_{3/2} h^2$ on the reheating temperature for different gravitino and messenger mass values and  $m_{\tilde{g}}=3$ TeV. In the right panel the abundance of the $10$ MeV gravitino reaches its thermal value. Evidently, the gravitino becomes overabundant once messengers get thermalized. }}
\end{figure}

When the reheating temperature of the universe is larger than $M_\text{mess}$ the messenger fields thermalize. The production of $\pm1/2$ helicity gravitinos takes places dominantly from messenger thermal scatterings whereas the production from MSSM scatterings is suppressed by $M^2/T^2$ compared to the low temperature $T<M_\text{mess}$ result. The relevant scattering cross section of a messenger particle $j$ with a MSSM particle $i$ to a Goldstino $\psi_G$ plus a messenger particle $j'$ is \cite{Choi:1999xm}
\begin{equation} \label{cs1}
\sum_{i, \, j,\, j'}  
\sigma(i+j   \rightarrow j' + \psi_G)  = \xi\, \lambda^2_\text{mess}(\bar{s}) \, f_{ij}(\bar{s}, M_\text{mess})\,,
\end{equation}
where $\bar{s}$ is the center of mass energy squared. The above process gives the leading contribution (see appendix B) to the scattering yield that reads
\begin{align}
Y^\text{mess(sc)}_{3/2}
\begin{split}
& = \int^{T_\text{rh}}_{T} \frac{\left\langle \sigma_{i+j\rightarrow j'+\psi_G} \, v_{ij} \, n_i\,  n_j \right\rangle}{\varsigma(T)H(T)T} dT \\
& = \,\frac{\bar{g}M_\text{Pl} \xi}{16\pi^4}\int^\infty_{x_\text{rh}}dx \, x^3 K_1(x) \int^{xT_\text{rh}}_{M_\text{mess}} d\sqrt{\bar{s}}\, \left\langle  \lambda^2_\text{mess}(\bar{s}) \right\rangle \,f_{ij}(\bar{s}, M_\text{mess}) \frac{\left[\bar{s}-M^2_\text{mess} \right]^2}{\bar{s}^2}\,,
\end{split}
\end{align}
where $x=M_\text{mess}/T$, $\bar{g}=135\sqrt{10}/(2\pi^3 g_*^{3/2}(T))$, $\varsigma(T)$ the entropy density, $g_*(T)$ the total number of the effectively massless degrees of freedom above the messenger scale,
$\xi=\alpha_1+3\alpha_2+8\alpha_3$, $K_1(x)$ a modified Bessel function and the yield $Y_{3/2}$ is the gravitino number density in units of entropy density $\varsigma(T)$. 
Writing the explicit form of the function $f_{ij}(\bar{s}, M_\text{mess})$ in terms of the temperature, $T=\sqrt{\bar{s}}$, we obtain 
\begin{equation}\label{YT}
\begin{split}
Y^\text{mess(sc)}_{3/2}=\frac{\bar{g}M_\text{Pl} \xi}{16\pi^4}\int^\infty_{x_\text{rh}}dx \, x^3 K_1(x) \int^{xT_\text{rh}}_{M_\text{mess}} dT\, 
&\left\langle \lambda^2_\text{mess}(T) \right\rangle \left(\frac{4}{T^2}-\frac{6M^2_\text{mess}}{T^4}+ \right.\\
& \left. +\frac{2M^4_\text{mess}}{T^6}+\frac{T^2-2M^2_\text{mess}}{T^4} \log \frac{T^4}{M^4_\text{mess}}\right)
\end{split}
\end{equation}

Turning now to the MSSM contribution, the gravitino yield in thermal scatterings  of the supersymmetric QCD plasma for $T'\ll M_\text{mess}, \, T_\text{rh}$ is 
\begin{equation}
Y^\text{MSSM(sc)}_{3/2} \approx \int^{M_\text{mess}}_{T'} dT \frac{\gamma_\text{sc}}{\varsigma(T)H(T)T}\,.
\end{equation}
The relevant gravitino production rate $\gamma_\text{sc}$ is found to be $\gamma_\text{sc} \sim 10^{-2}\, T^6 m^2_{\tilde{g}}/m^2_{3/2} M^2_\text{Pl}$ \cite{Bolz:2000fu}. 

In the following we consider the cases where the value of the messenger coupling $\lambda_\text{mess}$ is independent of the temperature, subsection (2.1), temperature dependent, subsection (2.2), and the case where both the messenger coupling and the messenger mass depend on the temperature, subsection (2.3).


\subsection{Constant $\lambda_{\text{mess}\,0}$}

\subsubsection{The messenger sector}
Here we will derive the known formula (\ref{mess1}) in order the temperature effects, introduced in the next subsection, to be clear. The yield, that we name $Y^\text{mess(sc)}_{3/2} [0]$, from messenger thermal scatterings for a temperature independent coupling $\lambda_{\text{mess}\,0}$ is
\begin{equation} \label{Y0}
\begin{split}
Y^\text{mess(sc)}_{3/2} [0]
& =\frac{\bar{g}M_\text{Pl} \xi}{16\pi^4} \,\frac{\lambda^2_{0\, \text{mess}}}{M_\text{mess}} \int^\infty_{x_\text{rh}}dx \,
 K_1(x) \left[\frac{248}{45} x^3-8x_\text{rh}x^2+ \right. \\
&\quad\quad\quad\quad\quad\quad \left. \frac{26}{9}x^3_\text{rh} -\frac{2}{5}\frac{x^5_\text{rh}}{x^2} -4(x_\text{rh}x^2-\frac{2}{3}x^3_\text{rh})\log\left(\frac{x}{x_\text{rh}} \right)  \right] \\
& \simeq \, \frac{3\pi}{2} \,\frac{248}{45}\, \frac{\bar{g} \xi}{16\pi^4} \,\frac{M_\text{Pl}}{M_\text{mess}}\,\lambda^2_{\text{mess}\,0}\,.
\end{split}
\end{equation}
The coupling $\lambda_{\text{mess}\,0}$ can be replaced by inserting the zero temperature relation (\ref{soft}) for the gaugino masses,
$\lambda^2_{\text{mess}\,0} = \left({4\pi}/{\alpha}\right)^2 \times ({m^2_{\tilde{g}}\, M^2_\text{mess}})/(3\, M^2_\text{Pl}\, m^2_{3/2})$,
and using the relation between the relic abundance and the yield for the gravitinos,
$\Omega^{\text{mess(sc)}}_{3/2}h^2 \sim \,2.8 \times 10^8\, Y^\text{mess(sc)}_{3/2} (m_{3/2}/\text{GeV})$,
 one obtains the expression (\ref{mess1}) 
\begin{equation}
\left. \Omega^\text{mess(sc)}_{3/2} h^2 \right|_{\lambda_{\text{mess}\,0}}    \sim \,  0.4\, \left(\frac{M_\text{mess}}{10^4 \,\text{GeV}}\right) \left(\frac{\text{GeV}}{m_{3/2}} \right) \left(\frac{m_{\tilde{g}}}{\text{TeV}}\right)^2\left[\frac{270}{g_*(T_\text{rh})}\right]^{3/2} \,.
\end{equation}

\subsubsection{The MSSM sector}
The production of Goldstinos from the thermalized MSSM plasma becomes efficient for temperatures below the messenger mass. 
In the conventional case where the messenger coupling is temperature independent, $\lambda_{\text{mess}\,0}$, the gravitino yield from the MSSM, $Y^{\text{MSSM(sc)}}_{3/2}[0]$, reads
\begin{equation} \label{Ynt}
 Y^{\text{MSSM(sc)}}_{3/2}[0] \,\approx \,10^{-22}\, \text{GeV}^{-1}\, \frac{m^2_{\tilde{g}}(T=0)}{12 m^2_{3/2}} \left[ \frac{228.75}{g_*(M_\text{mess})}\right]^{3/2}  \int^{M_\text{mess}}_{T'} \,dT\,.
\end{equation}
The yield is independent of the $\lambda_\text{mess}$ and it is dominated by temperatures $T\sim M_\text{mess}$. 
The corresponding relic abundance is given by the formula (\ref{mssm}) and depends on the scale of the messenger mass, $M_\text{mess}$, not on the reheating temperature.


\subsection{Temperature dependent $\lambda_\text{mess}(T)$}

We consider a model for the messenger sector described by the superpotential 
\begin{equation} \label{W1}
W_\text{mess} =\left(\frac{S}{\Lambda_*}\right)^n X \phi \bar{\phi} +M_\text{mess} \phi\bar{\phi} \,
\end{equation}
where the messenger coupling is controlled by the VEV of an extra singlet $S$, $\lambda_\text{mess}\equiv (S/\Lambda_*)^n$. The $X$ is the supersymmetry breaking spurion and we assume that the singlet $S$ is stabilized at non-zero VEV $S_0$ by some unspecified dynamics $\delta W_\text{stab}(S)$ 
on which we will comment in the next section. We consider that the $S$ is coupled to thermalized degrees of freedom via a Yukawa coupling $\kappa$. At zero temperature the VEVs of the $S$, $X$ and $\bar{\phi}\phi$ fields are 
\begin{equation}
S_0, \quad X_0=0\,, \quad  \phi\bar{\phi} = 0\,. 
\end{equation}

At finite temperature the effective potential for the fields can be much different than the zero temperature one altering the effective values of masses and couplings.
The temperature dependent 1-loop effective potential is of the form \cite{Kapusta:2006pm, Quiros:1999jp, Mukhanov:2005sc} 
\begin{equation}
V_1(\phi, T) = \frac{T^4}{2\pi^2} \left[\sum_{i} n_iJ_B\left(\frac{m^2_i(\varphi)}{T^2}\right)+\sum_{j}n_j J_F\left(\frac{m^2_j(\varphi)}{T^2}\right)  \right]
\end{equation}
where the functions $J_B$ and $J_F$ are defined by 
 $J_{F, B}=\int_0^\infty dy \, y^2 \log \left[1\pm e^{-\sqrt{y^2+M^2_i/T^2}} \right]$  
and $M^2_i$ is an eigenvalue of the mass squared matrices. In the high temperature limit where $T$ is much greater than the mass eigenvalues the temperature dependent scalar potential reads $V^{T}_1 \propto \sum_i m^2_i(\varphi)T^2$ plus terms linear to temperature that we omit here, an approximation sufficient for our scope.
We are interested in the way the thermal effects change the shape of the potential for the $S$ field and how the high temperature minimum evolves as the temperature decreases. We treat the spurion $X$ as a background field stabilized  thermally at the origin of the field space. This is a reasonable assumption because for temperatures larger than the messenger mass scale the $X$ and $S$ receive a non-zero thermal mass \cite{Kolb:1990vq} from the messenger fields that can drag them to the origin even though initially both fields can can be displaced. Small values for $\lambda_\text{mess}$ are also welcome for the selection of the SUSY breaking vacuum. Thus, the initial conditions we consider are $\left\langle X(T)\right\rangle \sim \left\langle S(T)\right\rangle_T \sim 0$ for $T > M_\text{mess}$. Details about the thermal evolution of the spurion $X$ can be found in ref. \cite{Dalianis:2010yk, Dalianis:2010pq}. 
By tracing the evolution of the singlet $S$ we can determine how the value of the messenger coupling $\lambda(S)$ changes with the temperature.

During the thermal phase the $S$ field receives a plasma mass $\sim \kappa T$ from thermalized degrees of freedom. 
Assuming a general polynomial expansion for the singlet potential at zero temperature, the relevant effective potential at finite temperature  reads approximately 
\begin{equation} \label{potT}
V(S,T) \sim   a (S+S^\dagger)+m^2_S |S|^2+ \kappa^2 T^2 |S|^2+ ...
\end{equation}
where the ellipsis is for terms that do not affect the evolution of the $S$, and we have assumed a typical stabilization mechanism for $S$
that generates the dimensionful parameters $a$, $m_S$.  The minimum for the $S$ field is temperature dependent and from (\ref{potT}) we obtain
\begin{equation} \label{ST}
\left\langle S(T) \right\rangle \sim \frac{a}{m^2_S+\kappa^2T^2}\, = \,S_0 \frac{1}{1+\left(\frac{\kappa T}{m_S}\right)^2}\,.
\end{equation}
Let us now define the temperature 
\begin{equation}
T_S\equiv \frac{m_S}{\kappa}
\end{equation}
 characteristic of the singlet transition to the zero temperature VEV $S_0$ \footnote{In case $a=0$ and negative $m^2_S$, $T_S=|m_S|/\kappa$ is the critical temperature of a second order phase transition from $S=0$ to $S=S_0$.}. If the singlet is tightly coupled to the plasma, with interaction rate $\Gamma_S$ larger that the Hubble expansion rate $H(T)$ \cite{Kolb:1990vq}, then the temperature fluctuations of $S$ are unsuppressed, $\delta S_T \simeq T/\sqrt{24}$, and have to be added to the thermal average value \cite{Mukhanov:2005sc}
\begin{equation}
S(T) =\left\langle S(T)  \right\rangle \pm \delta S_T\,.
\end{equation}
As the temperature decreases the scalar $S$ field moves towards the zero temperature minimum and the messenger coupling changes accordingly 
\begin{equation} \label{lT}
\lambda_\text{mess}(T) = \left[\frac{S(T)}{\Lambda_*}\right]^n\,.
\end{equation} 
For $n=1,\,2$ we attain
\begin{equation}
\begin{split}
&n=1, \quad\quad \left\langle \lambda^2_\text{mess}(T) \right\rangle = \frac{1}{\Lambda^{2}_*}  \left[ \left\langle S(T)  \right\rangle^2 +\delta S_T^2 \right] \\
&n=2, \quad\quad \left\langle \lambda^2_\text{mess}(T) \right\rangle = \frac{1}{\Lambda^{4}_*}  \left[ \left\langle S(T)  \right\rangle^4 +\delta S_T^4 + 6\left\langle S(T)  \right\rangle^2\delta S_T^2 \right]
\end{split}
\end{equation}
and in a compact form for arbitrary $n$
\begin{equation} \label{AV}
\left\langle \lambda^2_\text{mess}(T) \right\rangle = \frac{1}{\Lambda^{2n}_*} \left\langle \left[ \left\langle S(T)  \right\rangle \pm \delta S_T \right]^{2n} \right\rangle \,=\, \frac{1}{\Lambda^{2n}_*} \left\langle  \sum^{2n}_{k=0} 
\left(\begin{array} [pos]{c} 2n \\ k \end{array}\right) \left\langle S(T)  \right\rangle^{2n-k} \left( \pm \delta S_T \right)^{k} \right\rangle
\end{equation}
where we used the binomial identity. 

The interaction rate of the singlet with the thermalized degrees of freedom is  generally $\Gamma_S=\left\langle n_\text{eq} \sigma v \right\rangle \propto T$, whereas $H(T)\propto T^2$. This dependence on the temperature implies that the $\Gamma_S$ catches up the expansion rate as the temperature drops down.  We call $T_\kappa$ the temperature that the singlet joins the thermal equilibrium, that is 
\begin{equation}
\left. \Gamma_S = H(T)\right|_{T=T_\kappa}.
\end{equation}
The thermal dispersion of $\delta S_T$ can be suppressed if the singlet is out of the thermal equilibrium. To this end, there are two cases to be examined
$$
\sqrt{\left\langle \delta S^2_T  \right\rangle }  \, \sim \,
\begin{cases} 
  \frac{T}{\sqrt{24}}\,, \quad &  \Gamma_S > H(T) \quad \text{or} \quad T<T_\kappa\,, \quad \text{``thermal singlet"} \\ 
 0\,, \quad   & \Gamma_S < H(T) \quad  \text{or} \quad T>T_\kappa\,, \quad \text{``non-thermal singlet"}\,.
\end{cases}
$$  
Therefore, there are two pivotal temperatures concerning the thermal dynamics of the singlet: the $T_S$ for the $\left\langle S(T) \right\rangle$ evolution and the $T_\kappa$ for the thermal dispersion $\delta S_T$.

\subsubsection{The messenger sector}

The Goldstino yield from thermal scatterings of messenger fields for $M_\text{mess}<T_\kappa <T_\text{rh}$ reads
\begin{equation}\label{YT2}
\begin{split}
Y^\text{mess(sc)}_{3/2}= \frac{\bar{g}M_\text{Pl} \xi}{16\pi^4}\int^\infty_{x_\text{rh}}dx \, x^3 K_1(x) \times & \left[ \int^{xT_\kappa}_{M_\text{mess}} dT
\left\langle \left[ \frac{\left\langle  S(T) \right\rangle}{\Lambda_*} \pm  \frac{T}{\Lambda_* \sqrt{24}} \right]^{2n} \right\rangle \left(\frac{4}{T^2}-... \right) \right. \\
& \left. + \int^{xT_\text{rh}}_{xT_\kappa} dT  \left[\frac{\left\langle S(T) \right\rangle}{\Lambda_*} \right]^{2n} \left(\frac{4}{T^2}-... \right) \right]\,.
\end{split}
\end{equation}
We note that for $n=0$ the integral is dominated by the $M_\text{mess}$ value  whereas for $n>1$ by the reheating temperature. 
For $T_\kappa>T_\text{rh}>M_\text{mess}$  we get the thermal singlet case whereas for $T_\kappa<M_\text{mess}<T_\text{rh}$ we get the non-thermal singlet, that we discuss in detail below. 
\\
\\
\textbf{Thermal singlet} 
\\
At high temperatures the minimum of the effective potential for $S$ is about the origin, $\left\langle S(T)  \right\rangle \rightarrow 0$, and when $\Gamma_S >H(T)$ the thermally averaged messenger coupling squared (\ref{lT}) is approximately
\begin{equation} \label{therS}
\left\langle \lambda_\text{mess}^2(T) \right\rangle  =  \frac{1}{\Lambda^{2n}_*} \left\langle \left[ \left\langle S(T)  \right\rangle \pm \delta S_T \right]^{2n} \right\rangle \sim   \frac{1}{\Lambda^{2n}_*} \, \left(\frac{T}{\sqrt{24}} \right)^{2n}\,.
\end{equation}
For $T_\kappa>T_\text{rh}$ the yield (\ref{YT}) is given by the integral
\begin{equation} \label{In}
\begin{split}
\left. Y^\text{mess(sc)}_{{3/2}}\right|_{\delta S_T}\, \simeq \,
& \frac{\bar{g}\xi M_\text{Pl} }{16\pi^4}\frac{1}{(\Lambda_*\sqrt{24})^{2n}} \int^\infty_{x_\text{rh}}dx \, x^3  K_1(x) \times \\
&  \frac{4}{2n-1} (xT_\text{rh})^{2n-1} \left[1+ \frac{1}{2n-1}\left(-1+(2n-1)\log x\, +\, (2n-1) \log\frac{T_\text{rh}}{M_\text{mess}}  \right)\right]+ ...
\end{split}
\end{equation}
for $n\geq 1$ and the ellipsis corresponds to subleading terms. If it is $T_\kappa<T_\text{rh}$ the $T_\text{rh}$ has to be replaced by the $T_\kappa$ temperature. For $n=1$, that is for messenger superpotential $W_\text{mess} =(S/\Lambda_*) X\phi\bar{\phi}$, the yield is
\begin{equation}
\left. Y^\text{mess(sc)}_{{3/2}} \right|_{\delta S_T} \, \sim \, \frac{\bar{g} \xi M_\text{Pl} }{16\pi^4} \frac{T_\text{rh}}{(\Lambda_*\sqrt{24})^{2}} \left[87.6 + 64 \log\frac{T_\text{rh}}{M_\text{mess}}\right]\,, \quad\quad n=1
\end{equation} 
while for $W_\text{mess} =(S/\Lambda_*)^2 X\phi\bar{\phi}$, i.e. $n=2$, we take
\begin{equation}
\left. Y^\text{mess(sc)}_{{3/2}}\right|_{\delta S_T}\, \sim \, \frac{\bar{g}\xi M_\text{Pl}}{16\pi^4} \frac{T^3_\text{rh}}{(\Lambda_*\sqrt{24})^{4}} \left[1255 + 512 \log\frac{T_\text{rh}}{M_\text{mess}}\right]\,, \quad\quad n=2\,.
\end{equation} 
These results should be compared with the yield (\ref{Y0}),  $Y^\text{mess(sc)}_{{3/2}}[0] \simeq 26  \lambda^2_{\text{mess}\,0}\,\bar{g}\xi M_\text{Pl}/({16\pi^4}M_\text{mess})$, where the messenger coupling is constant. We see that the Goldstino yield due to thermal scatterings of messengers (\ref{In}) for thermally excited $S$ and $\left\langle S(T)\right\rangle \rightarrow 0$ is weaker than the conventional result (\ref{mess1})  when 
\begin{equation} \label{ampl}
S_0 > \left(M_\text{mess} \, T_\text{rh}^{2n-1} \right)^{\frac{1}{2n}}\,
\end{equation}
and amplified for smaller $S_0$ values.
The relic abundance of $\pm1/2$ helicity gravitinos from messenger scatterings is thereby
\begin{equation} \label{OdS}
\left. \Omega^{\text{mess(sc)}}_{3/2}h^2 \right|_{\delta S_T}\sim\, 0.1 \left(\frac{T^{2n-1}_\text{rh}M_\text{mess}}{S^{2n}_0} \right)  \,\left(\frac{\text{GeV}}{m_{3/2}} \right) \left(\frac{ M_\text{mess} }{10^4\, \text{GeV}} \right) \left(\frac{m_{\tilde{g}}}{\text{TeV}}\right)^2.
\end{equation}
Apparently, the ratio $S_0 / \left(M_\text{mess} \, T_\text{rh}^{2n-1} \right)^{\frac{1}{2n}}$ rules the impact of the singlet thermal dispersion on the gravitino production rate from the messenger fields. 
\\
\\
\textbf{Non-thermal singlet}
\\
For  $\Gamma_S < H(T)$ or equivalently $T_\kappa<M_\text{mess} < T_\text{rh}$ the singlet is decoupled and its abundance is non-thermal. Nevertheless, the singlet receives a plasma mass $\kappa T$. In such a case it is $\delta S_T \ll T/\sqrt{24}$ and thus $S(T) \simeq \left\langle S(T) \right\rangle$. The thermally averaged messenger coupling squared is 
\begin{equation} \label{lamT}
\left\langle \lambda^2_\text{mess}(T) \right\rangle \sim \left[\frac{\left\langle S(T) \right\rangle}{ \Lambda_*} \right]^{2n} =  \frac{\lambda^2_{\text{mess}}(0)}{\left[1+\left(\frac{T}{T_S}\right)^2\right]^{2n}}
\end{equation}
where  the thermal evolution (\ref{ST}) of the singlet $S$ was considered. 
For $n\geq 1$ the yield reads
\begin{equation}\label{YTn}
\begin{split}
\left. Y^\text{mess(sc)}_{3/2}\right|_{\left\langle S(T) \right\rangle}= \frac{\bar{g}M_\text{Pl} \xi}{16\pi^4}  & \int^\infty_{x_\text{rh}}dx \, x^3 
 K_1(x) \int^{xT_\text{rh}}_{M_\text{mess}} 
 dT\, \left[\lambda_{\text{mess}}(0)\right]^{2} {\left[1+\left(\frac{T}{T_S}\right)^2\right]^{-2n}} \times \\
& \left(\frac{4}{T^2}-\frac{6M^2_\text{mess}}{T^4} +\frac{2M^4_\text{mess}}{T^6}+\frac{T^2-2M^2_\text{mess}}{T^4} \log \frac{T^4}{M^4_\text{mess}}\right)\,.
\end{split}
\end{equation}
For $n=1$, we numerically find that the integral gives a 65 times suppressed yield for $T_S/M_\text{mess}=1$, while for $T_S/M_\text{mess}\sim 0.1$ the suppression is $10^{5}$ times! The suppression of the Goldstino yield, for the temperature dependent coupling (\ref{lT}) and non-thermal singlet, is depicted in Figure 4. For higher values of $n$ the suppression is even more drastic. 
Therefore, the helicity $\pm1/2$ gravitino abundance from messenger thermal scatterings for $T_S<M_\text{mess}$ is $\epsilon^{-1}$ times suppressed
\begin{equation} \label{Oepsilon}
\left. \Omega^{\text{mess(sc)}}_{3/2} h^2 \right|_{\left\langle S(T) \right\rangle}  \sim \epsilon \, \,\left(\frac{\text{GeV}}{m_{3/2}} \right) \left(\frac{ M_\text{mess} }{10^4\, \text{GeV}} \right) \left(\frac{m_{\tilde{g}}}{\text{TeV}}\right)^2.
\end{equation}
where $\epsilon\equiv Y^\text{mess(sc)}_{3/2}/ Y^\text{mess(sc)}_{3/2}[0]\ll 1$. For $T_S\ll M_\text{mess}$ the above relic abundance is practically zero. Note that for a second order phase transition ($a=0$, $m^2_S<0$)  $\epsilon=0$ for $T_S=|m_S|/\kappa <M_\text{mess}$.


\begin{figure} 
\centering
\includegraphics [scale=.8, angle=0]{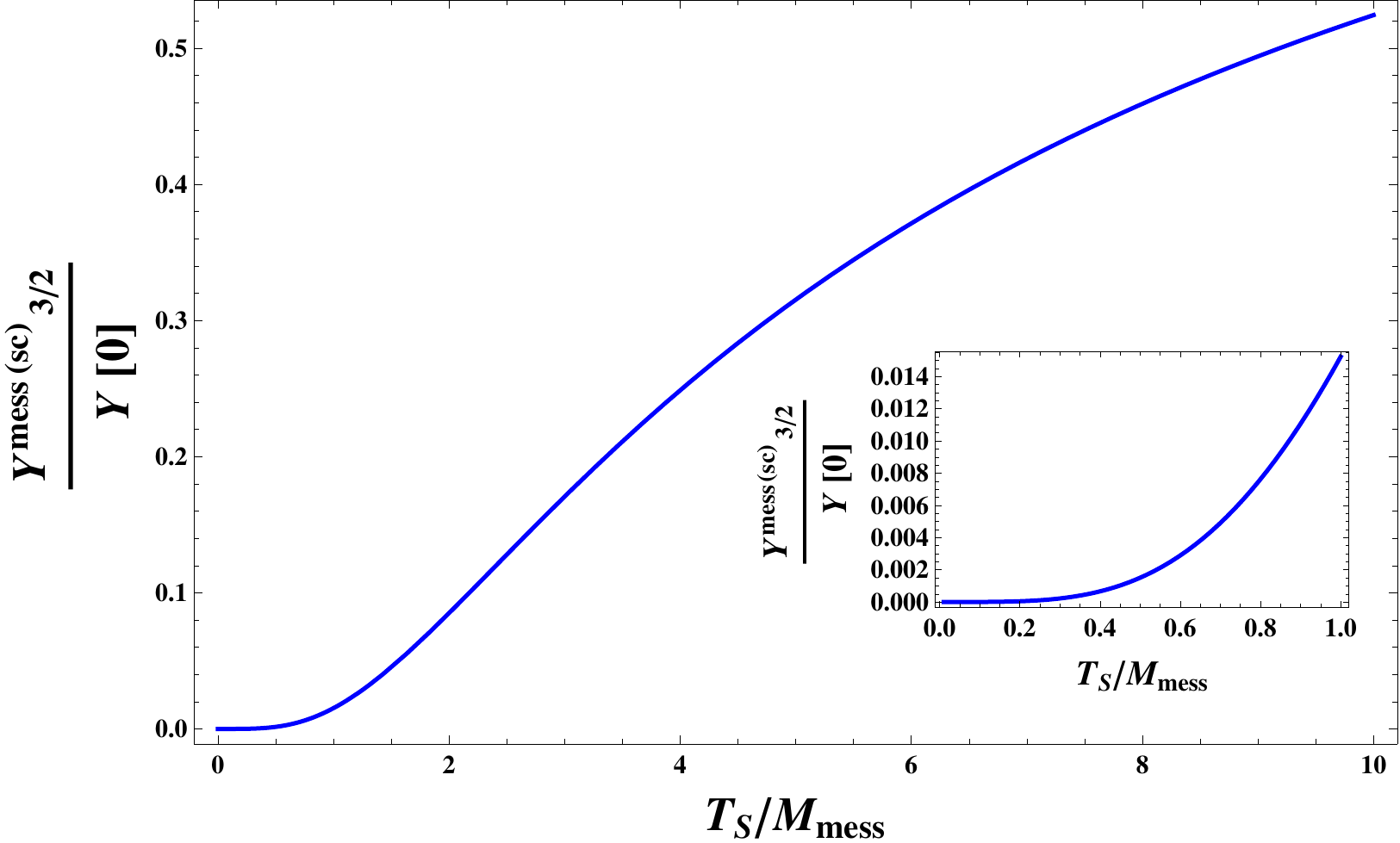} 
\caption {\small {The suppression of the Goldstino yield from messenger fields thermal scatterings due to a temperature dependent coupling $\lambda_\text{mess} = \left\langle S(T) \right\rangle/\Lambda_*$ is depicted. The smaller plot is a zoom in the origin of the big plot. We see that for $T_S \ll M_\text{mess}$ the yield is vanishing. The $Y{[0]}$ stands for the Goldstino yield from thermal scatterings for constant messenger coupling $\lambda_{\text{mess}\,0}$.}}
\end{figure}

\subsubsection{The MSSM sector}
When the temperature of the Universe drops below the $M_\text{mess}$ the gravitino production rate from the MSSM sector increases.
The MSSM contribution is crucial since the thermal production of helicity $\pm1/2$ gravitinos from messengers can be safely neglected for $T_S \ll M_\text{mess}$ and non-thermal singlet.  
Due to dynamic nature of the messenger coupling the gaugino masses are temperature dependent (for $M_\text{mess}\neq 0$)
\begin{equation} \label{suppr}
m_{\tilde{g}_i} (T) = \frac{\alpha_i}{4\pi} \frac{\lambda_\text{mess}(T)\, F}{M_\text{mess}} \,= \, m_{\tilde{g}_i}(T=0)\, \frac{\lambda_\text{mess}(T)}{\lambda_{\text{mess}}(0)}\,
\end{equation}
where $\lambda_\text{mess}(T=0)=(S_0/\Lambda_*)^n$. 
Thereby, the gravitino yield from the thermalized MSSM is given by the expression
\begin{equation} \label{gravMSSM}
\begin{split}
 Y^{\text{MSSM(sc)}}_{3/2} \, & \approx \,10^{-22}\, \text{GeV}^{-1}\, \frac{m^2_{\tilde{g}}(T=0)}{12 m^2_{3/2}} \left[ \frac{228.75}{g_*(M_\text{mess})}\right]^{3/2} \int^{M_\text{mess}}_{T'} \,dT\, \frac{\left\langle \lambda^2_\text{mess}(T)\right\rangle}{\lambda^2_\text{mess}(0)} \\
& \approx \,  \frac{Y^{\text{MSSM(sc)}}_{3/2}[0]}{M_\text{mess}}  \int^{M_\text{mess}}_{T'} \,dT\, \frac{\left\langle \left[ \left\langle S(T)  \right\rangle \pm \delta S_T\right]^{2n}\right\rangle}{S_0^{2n}} \,.
\end{split}
\end{equation}
where the result for the conventional yield (\ref{Ynt}) was used. Plugging in the thermal evolution of the singlet's VEV (\ref{ST}) and the considering unsuppressed thermal dispersion $\delta S_T=T/\sqrt{24}$ the yield is recast into
\begin{equation}
\begin{split}
Y^{\text{MSSM(sc)}}_{3/2} \,  \approx \, &  \frac{Y^{\text{MSSM(sc)}}_{3/2}[0]}{M_\text{mess}}   \int^{M_\text{mess}}_{T_\kappa} \,dT\, \left[{ 1+\left(\frac{T}{T_S}\right)^2} \right]^{-2n}\\
& + \frac{Y^{\text{MSSM(sc)}}_{3/2}[0]}{M_\text{mess}} \int^{T_\kappa}_{T'} \,dT\, \left\langle\left[\frac{1}{ 1+\left(\frac{T}{T_S}\right)^2} \pm \frac{T}{ S_0 \sqrt{24}}\right]^{2n}\right\rangle  \,.
\end{split}
\end{equation}
\\
\\
\textbf{Thermal singlet}
\\
If the singlet is  thermalized and trapped in the origin of the field space then the messenger coupling is given by the thermal dispersion of the singlet (\ref{therS}), $\lambda_\text{mess}(T) \sim (\delta S_T/\Lambda_*)^n$.  For $T_S<T_\kappa<M_\text{mess}<T_\text{rh}$  the gravitino yield reads
\begin{equation} \label{Yt}
\left.  Y^{\text{MSSM(sc)}}_{3/2}\right|_{\delta S_T} \,\sim \,10^{-22}\, \text{GeV}^{-1}\, \frac{m^2_{\tilde{g}}(T=0)}{12 m^2_{3/2}} \left[ \frac{228.75}{g_*(T_m)}\right]^{3/2}  \int^{T_\kappa}_{T'} \,dT\, \left(\frac{T}{S_0\sqrt{24}}\right)^{2n}  \,
\end{equation}
and $\delta S_T=T/\sqrt{24}$. The relic gravitino energy density is given by the expression
\begin{equation} \label{OTMSSM}
\left. \Omega^{\text{MSSM(sc)}}_{3/2}h^2\right|_{\delta S_T} \sim \frac{0.1}{(2n+1)(24)^n} \left( \frac{T_\kappa}{S_0} \right)^{2n}   \left( \frac{T_\kappa}{10^8 \,\text{GeV}}\right) \left(\frac{\text{GeV}}{m_{3/2}} \right) \left(\frac{m_{\tilde{g}}}{\text{TeV}}\right)^2 \,.
\end{equation}
We see that the thermal dispersion of the singlet does not amplify the gravitino production rate, compared to the conventional yield (\ref{Ynt}), when 
\begin{equation}
S_0 \gtrsim \frac{T_\kappa}{\sqrt{24}} \left(\frac{T_\kappa}{ M_\text{mess}} \right)^{1/2n}\,.
\end{equation}
that is, for $S_0$ sufficiently large or $T_\kappa$ sufficiently small. 
\\
\\
\textbf{Non-thermal singlet}
\\
When the singlet is decoupled from the thermal plasma or its thermal dispersion is much suppressed then 
the value of the messenger coupling is determined by the temperature dependent VEV of the singlet (\ref{lamT}), $\lambda(T)=[\left\langle S(T) \right\rangle/\Lambda_*]^n$. The messenger Yukawa coupling, and so the gaugino masses, are suppressed for temperatures $T>T_S$. The relic gravitino yield from the thermalized MSSM plasma for $\lambda^2_\text{mess}(T)=\lambda^2_{\text{mess}}(0)\left[1+\left(T/T_S\right)^2\right]^{-2n}$ and $T_\text{rh}>M_\text{mess}$ reads
\begin{equation} \label{YntM}
 \left. Y^{\text{MSSM(sc)}}_{3/2}(T)\right|_{\left\langle S(T) \right\rangle} \sim 10^{-22}\, \text{GeV}^{-1} \, \left[ \frac{m^2_{\tilde{g}} (T=0)}{12 m^2_{3/2}} \right]\left[ \frac{228.75}{g_*(T)}\right]^{3/2} 
  \int^{M_\text{mess}}_{T'} dT\, \left[1+\left(\frac{T}{T_S}\right)^2\right]^{-2n}\,.
\end{equation}
For $n=1$ the integral in (\ref{YntM}) is
\begin{equation} \label{apI}
\left. \int^{M_\text{mess}}_{T'} dT\, \left[1+\left(\frac{T}{T_S}\right)^2\right]^{-2} = \frac{T_S}{2} \left\{\frac{T_S T}{T^2_S+{T}^2} +\text{Arctan}\left(\frac{T}{T_S}\right) \right\} \right|^{M_\text{mess}}_{T'} \,.
\end{equation}
For $M_\text{mess}> T_S$, the integral converges to $T_S \cdot\theta/2$;
where, $\theta \equiv \text{Arctan}(M_\text{mess}/T_S)$ a coefficient that takes values of order one, $\pi/4<\theta<\pi/2$.
Therefore, for $T_S < M_\text{mess} $ the (\ref{YntM}) is $ Y^{\text{MSSM(sc)}}_{3/2}(T) \propto T_S$.  The resulting gravitino abundance reads 
\begin{equation} \label{OgA}
\left. \Omega^{\text{MSSM(sc)}}_{3/2} h^2 \right|_{\left\langle S(T) \right\rangle}  \sim 0.1 \,\frac{\theta}{2} \, \left(\frac{ T_S}{10^8\,\text{GeV}} \right) \left(\frac{\text{GeV}}{m_{3/2}} \right) \left(\frac{m_{\tilde{g}}}{1\, \text{TeV}} \right)^2, \quad \text{for}\,\, T_\text{rh}>T_S\,.
\end{equation}
This expression is insensitive to the reheating temperature rather, it is sensitive only to the $T_S$ temperature and slightly to the messenger mass scale via the order one parameter $\theta$. 
 
The $\Omega^{\text{MSSM(sc)}}_{3/2}$ is the final result only if the thermalization of the singlet at $T_\kappa$ can be safely neglected. This is the case if the $T_\kappa$ is sufficiently small compared to $S_0$ or never comes (this occurs when the $\varphi, \bar{\varphi}$ fields decouple from the plasma before $S$ gets thermalized). 
The  condition 
\begin{equation} \label{S0M}
S_0 \, > \, \frac{T_\kappa}{\sqrt{24}}\left(\frac{T_\kappa}{T_S}\right)^{1/2n}
\end{equation}
ensures that the thermal dispersion of the singlet can be neglected.
Summarizing, for $T_S < M_\text{mess}$ and non-thermal singlet $S$, the total helicity $\pm1/2$ gravitino abundance produced from thermal scatterings is
\begin{equation}
\left. \Omega^{\text{mess(sc)}}_{3/2}+\Omega^{\text{MSSM(sc)}}_{3/2}\, \sim  \,\Omega^{\text{MSSM(sc)}}_{3/2}\right|_{\left\langle S(T) \right\rangle}\,. 
\end{equation}

\begin{figure} 
\centering
\includegraphics [scale=.9, angle=0]{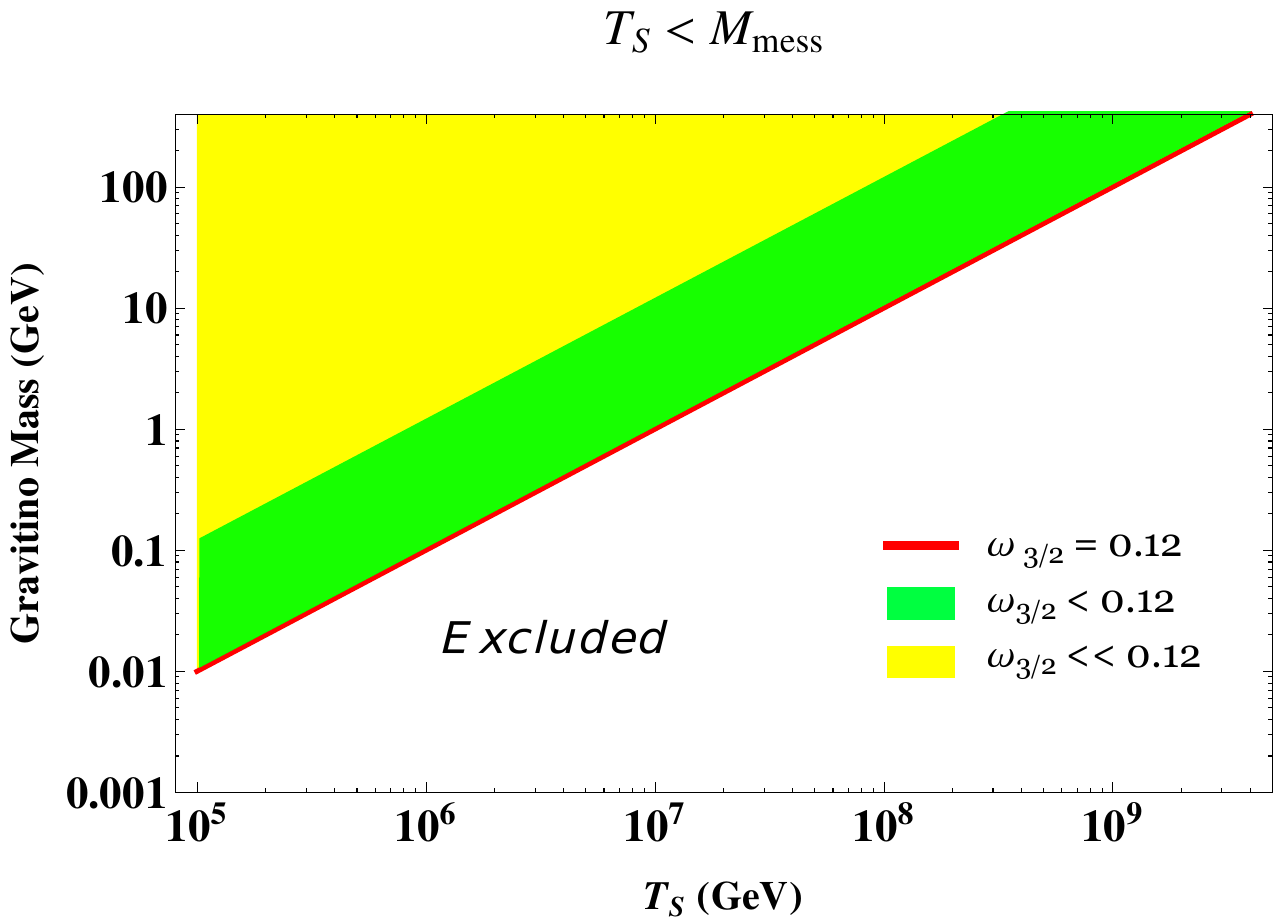} 
\caption {\small {The allowed regions of gravitino abundance $\omega_{3/2}\equiv \Omega_{3/2} h^2$ in the contour ($m_{3/2}, T_S$), for $m_{\tilde{g}}=3$ TeV, $T_S<M_\text{mess}$ and negligible thermal dispersion for the singlet. The reheating temperature can be larger than the messenger scale. 
}}
\end{figure}


\subsection{Zero bare mass $M_\text{mess}=0$}

When $M_\text{mess}=0$  the zero temperature VEVs of the $S$, $X$ and $\phi \bar{\phi}$ fields in the messenger superpotential (\ref{W1}) are 
\begin{equation}
S_0, \quad X_0 \neq 0\,, \quad  \phi\bar{\phi} = 0\,. 
\end{equation}
The messenger mass is determined by the VEVs of the background spurion field $X_0$ and the singlet $S_0$, and at zero temperature it is $\lambda_{\text{mess}}(0) X_0$. At finite temperature, the VEV of $S$ changes (the VEV of the spurion $X$ also changes however, the characteristic temperature that the $X$ settles into the $X_0$ vacuum is typically  $T_X>T_S$ \cite{Dalianis:2010pq}, though the $\lambda_\text{mess}(S)$ scenario calls for a dedicated analysis for this extended system of fields.) Hence, the messenger mass becomes temperature dependent $M_\text{mess}(T)=\lambda_\text{mess}(T) X_0$ and messengers fields become nonrelativistic at the temperature $T_\text{mess}$ defined by
\begin{equation} \label{Tmess}
\left. T =\lambda_\text{mess}(T) X_0 \,\right|_{T=T_\text{mess}}.
\end{equation}
For non-thermal singlet, the evolution of the VEV $\left\langle S(T) \right\rangle$ determines the evolution of the messenger coupling (\ref{lamT}) and it is $T_\text{mess}={\lambda_{\text{mess}}(0) \,X_0}\left[1+\left({T_\text{mess}}/{T_S}\right)^2\right]^n$. For $T_S \ll \lambda_{\text{mess}}(0) X_0$ we find that 
\begin{equation}
T_\text{mess} \, \sim \, \left[ T^{\,2n}_S \, \lambda_{\text{mess}}(0) X_0 \right]^{1/2n+1}.
\end{equation}
The relic abundance of Goldstinos from scattering of thermalized messengers is given by the expression (\ref{Oepsilon}). 
Again for $T_S < T_\text{mess}$ the coupling $\lambda_\text{mess}(T)$ is very suppressed and the Goldstino yield from messenger scatterings is negligible.

For temperatures below $T_\text{mess}$ the messengers are nonrelativistic.
Even though the messenger mass may still be thermally suppressed, $\lambda_\text{mess}(T) X_0 \ll \lambda_{\text{mess}} (0)X_0$, the supersymmetry breaking effects are mediated intact to the visible sector  once $T<T_\text{mess}$ because the soft masses are independent of the magnitude of the Yukawa coupling $\lambda_\text{mess}(T)$, given that $(\lambda_\text{mess} X_0)^2 > \lambda_\text{mess} F$.  
Indeed, from the formula (\ref{soft}) we can see that the $\lambda_\text{mess}$ drops out from the ratio $\lambda_\text{mess} F/ M_\text{mess}$, 
\begin{equation}
m_{\tilde{g}_i} = \frac{\alpha_i}{4\pi} \frac{ F}{X_0}\,
\end{equation}
and the gaugino fields are not modulated by thermal effects. 
The relic abundance of $\pm 1/2$ gravitinos from the MSSM depends here on the $T_\text{mess}$ and not on $T_S$ (see expression (\ref{OgA}))
\begin{equation} \label{OgB}
\Omega^{\text{MSSM(sc)}}_{3/2}h^2 \sim 0.1 \, \left(\frac{ T_\text{mess}}{10^8\,\text{GeV}} \right) \left(\frac{\text{GeV}}{m_{3/2}} \right) \left(\frac{m_{\tilde{g}}}{1\, \text{TeV}} \right)^2\quad \text{for} \,\,T_\text{rh}>T_\text{mess}\,.
\end{equation}
This is actually the conventional expression (\ref{mssm}) with the replacement $M_\text{mess} \rightarrow T_\text{mess}$.
The interesting feature here is that the thermal dispersion of the singlet does not amplify the gravitino production rate once the temperature falls below $T_\text{mess}$. This is due to the fact that for $M_\text{mess}=0$ the gaugino masses are independent of the $\lambda_\text{mess}$ value. Therefore, the condition 
\begin{equation}
T_\kappa < T_\text{mess}
\end{equation}
implies that the thermal dispersion of the singlet $\delta S_T$ can be totally neglected from the estimation of the gravitino yield. 


\section{Coupling the singlet $S$ to the thermal plasma}

Until know we have implicitly assumed that the singlet is coupled to some thermalized superfields $\varphi$ that we have not named. These fields may be exotics that participate in the equilibrium and couple to the singlet stabilizing it thermally at the origin.  An example is the superpotential interaction
\begin{equation} \label{wtab}
W= \left(\frac{S}{\Lambda_*}\right)^n X\phi\bar{\phi} +\kappa_\varphi S \varphi\bar{\varphi}\,.
\end{equation}

\begin{table}
\centering 
\begin{tabular}{|| l | p{0.70\textwidth} ||} 
\hline\hline 
$T_\text{rh}$ & the reheating temperature of the Universe
\\[0.5ex]
\hline
$T_S$ &   the $S$ transition to $S_0$ takes place (typically $T_S \sim m_S/\kappa$)
\\[0.5ex]
\hline
$T_\kappa$ &  the singlet $S$ joins the thermal equilibrium
\\[0.5ex]
\hline
$M_\text{mess}$ &  the messengers fields (with a bare mass) become non-relativistic 
\\[0.5ex]
\hline
$T_\text{mess}$ &  the messengers fields (with no bare mass) become non-relativistic
\\[0.5ex]
\hline
$T_X$ &  the VEV $X_0$ for the SUSY breaking spurion field appears
\\[0.5ex]
\hline
\hline 
\end{tabular}
\caption{Characteristic temperatures that are mentioned in the text.}  
\end{table}

\subsection{Cosmological constraints}
Assuming that the fields $\varphi$ interact with the thermal plasma via gauge fields with coupling $g^2_V\sim 1$ then the scattering cross section for a singlet production with $\varphi$ fields for processes of the form $S+1 \leftrightarrow 2+3$, which regulates the abundance of singlets,  is
\begin{equation}
\sigma(\bar{s}) \, \sim \,0.1 \frac{\kappa^2_\varphi}{\bar{s}}\,,
\end{equation}
where $\bar{s}$ the center of mass energy squared. It follows that the interaction rate reads
\begin{equation} \label{goldeq}
\Gamma_{S}= \left\langle n_\text{eq}\sigma v\right\rangle \sim \left\langle 0.1   \kappa^2_\varphi T^3/\bar{s} \right\rangle =  0.1 \, \kappa^2_\varphi T 
\end{equation}
where $n_\text{eq}=[\zeta(3)/4\pi] g\,T^3$ the equilibrium number density for $\varphi$ and  g their internal degrees of freedom, and $v\sim1$.
At very high temperatures the singlet interaction rate cannot follow the expansion rate until the temperature drops down at a specific value that we have named $T_\kappa$. It is 
\begin{equation}
T_\kappa \sim \frac{1}{{\cal O}(50)} \kappa^2_\varphi M_\text{Pl}\,.
\end{equation}
where $H(T)= 0.33 g^{1/2}_* T^2/M_\text{Pl}\sim 5 T^2/M_\text{Pl}$.
The out-of-equilibrium condition $\Gamma_{S} < H(T)$ at the messenger scale $M_\text{mess}$, or in other words $M_\text{mess}>T_\kappa$, implies that 
\begin{equation} \label{1}
\kappa_\varphi < \left[{\cal O}(50)\frac{M_\text{mess}}{M_\text{Pl}}\right]^{1/2}\,,
\end{equation}
This condition constrains the Yukawa coupling $\kappa_\varphi$ to very small values $\kappa_\varphi \ll 1$.

\begin{figure} 
\centering
\begin{tabular}{cc}
{(a)} \includegraphics [scale=.5, angle=0]{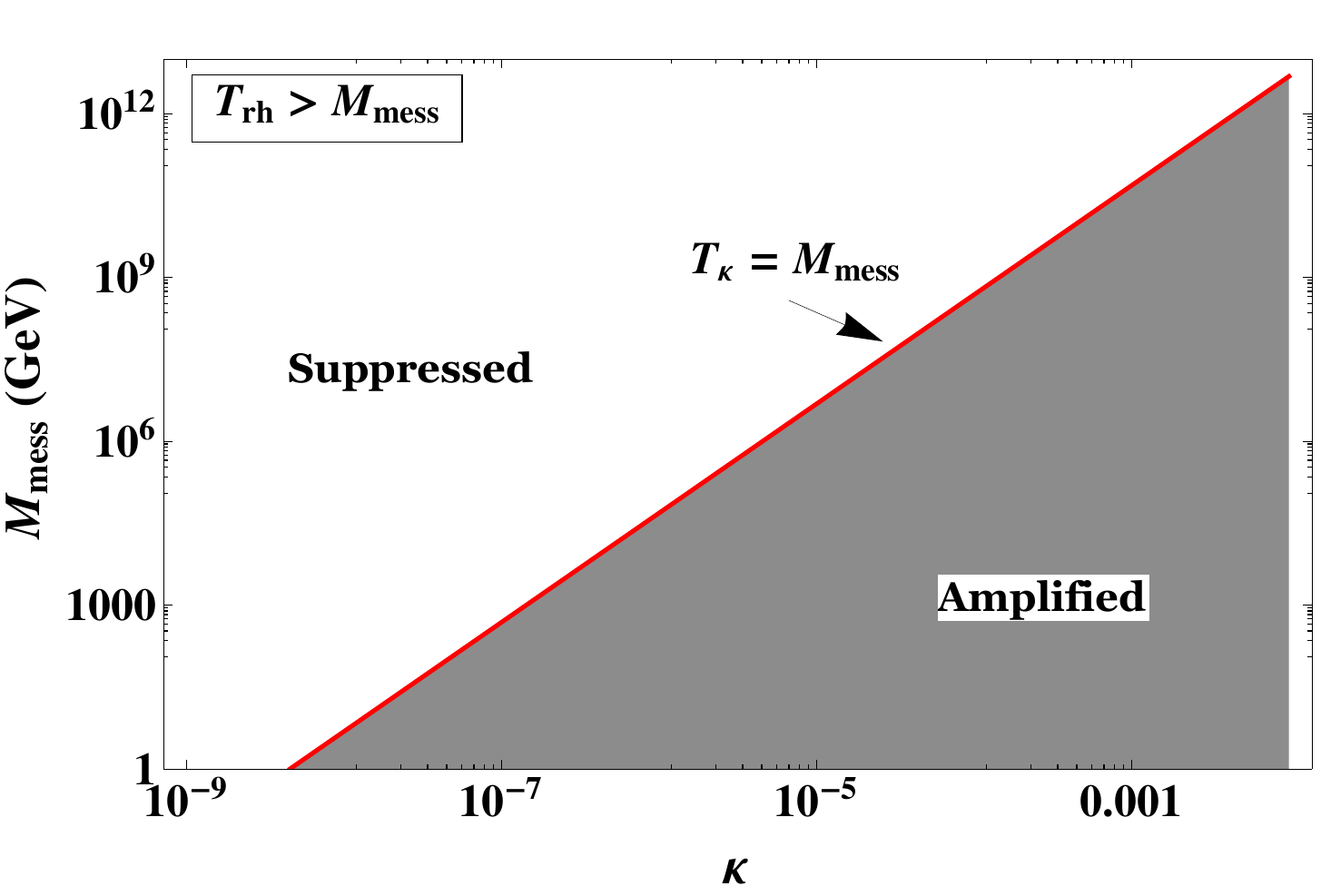} &
{(b)} \includegraphics [scale=.5, angle=0]{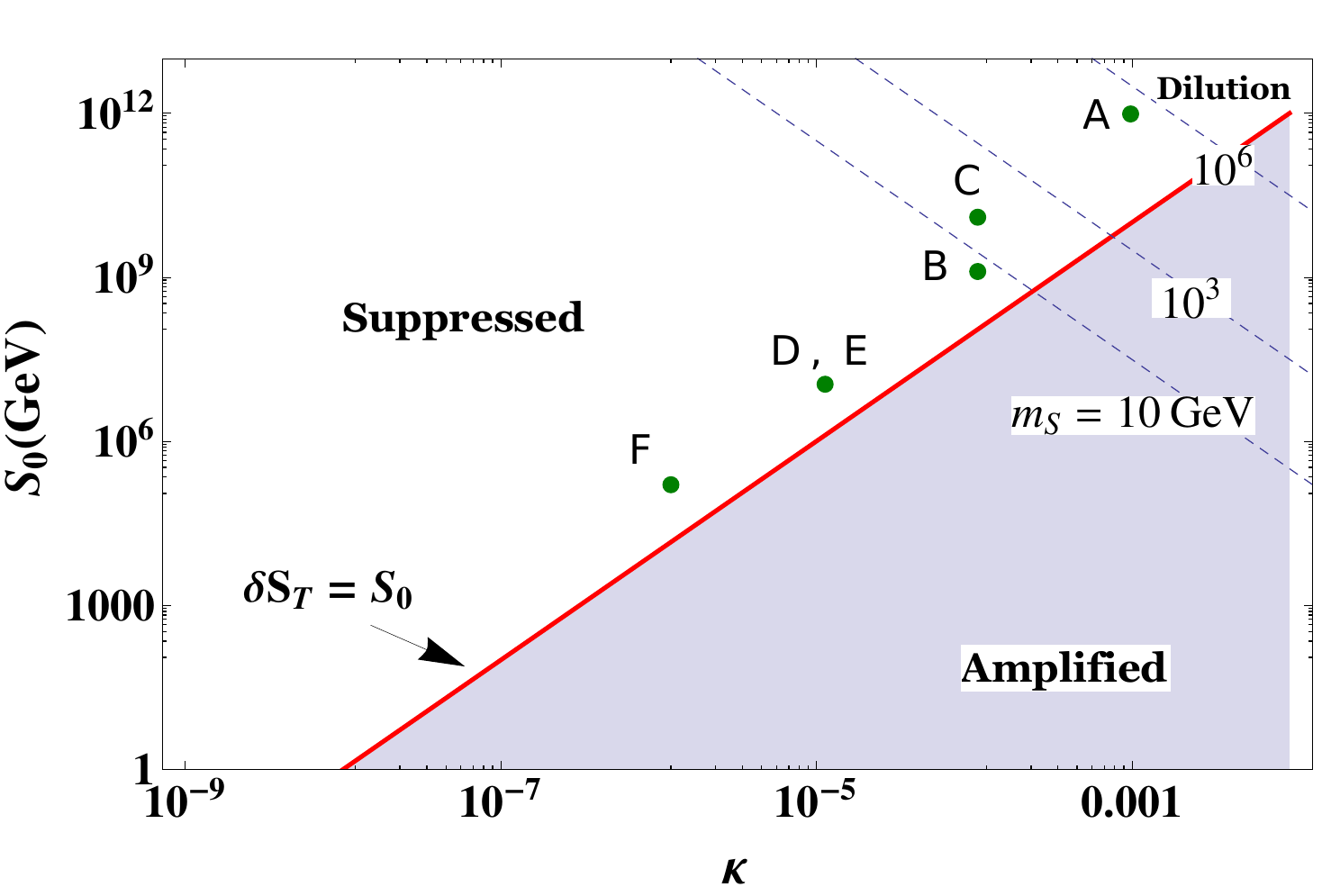}  \\
\end{tabular}
\caption{\small{ In the white part of the plots the relic gravitino abundance $\Omega_{3/2}$ is suppressed due to the suppression of the messenger coupling $\lambda_\text{mess}(T)$. In the dark part the $\lambda_\text{mess}(T)$ is effectively amplified due to the thermal dispersion of the singlet $S$ and there the $\Omega_{3/2}$ increases. The amplification is cosmologically more constrained in  the left panel, due to messengers significant contribution to $\Omega_{3/2}$. In the right panel, the amplification increases the $\Omega_{3/2}$ from MSSM scatterings; in the upper right corner over the dashed lines  thermal inflation may take place (for different $m_S$ values) because of the thermal trapping of $S$ in the origin. The $\kappa$ collectively stands for the coupling of the singlet to thermalized degrees of freedom. The points (A,B,...) correspond to the benchmark values of Table 2.}}
\end{figure}

On the other hand, the smallness of the coupling is welcome because it prevents $S$ from dominating the energy density of the early Universe. After inflation the Universe is reheated and the coupling of $S$ to the plasmon $\varphi$ and messenger fields generate a thermal potential for the singlet that $\delta V \sim \kappa^2_\varphi T^2|S|^2$ that shift the $S$ to the origin of the field space. The thermal trapping in the false vacuum $S=0$ may cause thermal inflation \cite{Lyth:1995ka} to take place. Thermal inflation can occur only for temperatures $T>T_S$ and for $V(S= 0)\sim m^2_S S^2_0 \gtrsim T^4$. Actually, the critical temperature $T_S=m_S/\kappa_\varphi$ is too large if the coupling is $\kappa_\varphi \ll 1$, even for small singlet mass values e.g. $m_S \sim 100$ GeV. One can see that for $\kappa^2_\varphi S_0<m_S$ thermal inflation cannot occur.  

At the temperature $T_S$ the minimum of the effective potential approximately coincides with the zero temperature VEV. It is actually $H(T_S)\sim T^2_S/M_P \ll m_S$ therefore the singlet follows the shift of the temperature dependent minimum and oscillations are damped. Otherwise, if the singlet starts oscillating about the zero temperature minimum with initial amplitude $S_0$, it will dominate the energy density of the universe only at the late temperature $T_\text{dom} \sim {m^2_S}S^2_0/ T_S^3$. Even in this case, the oscillations of the singlet do not cause any late entropy production if the singlet has a typical decay rate.

In addition to the smallness of the the Yukawa coupling to the thermalized degrees of freedom $\kappa_\varphi \ll 1$,  the VEV $S_0$ of the singlet has to be relatively large in order the gravitino abundance to be suppressed, see (\ref{S0M}), - except if $M_\text{mess}=0$. The first condition keeps the singlet outside the thermal equilibrium at high temperatures while the second condition prevents the Goldstino coupling to the MSSM  $\sim m^2_{\tilde{g}}/m^2_{3/2}$ from being amplified due to thermal dispersion $\delta S_T$ at lower temperatures. 

To recap, 
the relic gravitino abundance is given by the expression (\ref{OgA}), $\Omega_{3/2}h^2\propto T_S$, when $T_S<M_\text{mess}$ and $\kappa \ll 1$ and $S_0$ sufficiently large.

\subsection{Coupling the singlet $S$ to the Higgs bilinear}

An economical and attractive option is to employ the Higgs multiplets for the thermal stabilization of the $S$. 
This choice is also welcome for it could provide a solution to the $\mu/B\mu$ problem in GMSB. The common strategy to address the $\mu/B\mu$ problem is to introduce a symmetry such as the Peccei-Quinn (PQ) or the $Z_3$ symmetry that forbids the $\mu$-term in the exact symmetry limit. The $\mu$-term is replaced by the VEV of a weak-scale singlet, which is charged under this symmetry. The singlet $S$ can play this role with superpotential interactions with the Higgs bilinear as
\begin{equation} \label{pq1}
\delta {\cal L}  =   \int d^2 \theta\, \kappa_H \frac{S^a}{M^{a-1}_\text{Pl}}   H_u H_d +\text{h.c.}
\end{equation}
and $\mu=\kappa_H  \left\langle S\right\rangle^{a}/M^{a-1}_\text{Pl}$.

\begin{table}
\centering 
\begin{tabular}{||r|r|r|r|r|r|r||r|r||r||r||} 
\hline \hline
  & {$m_S$} &  $S_0$ & $\kappa_\varphi\,$ & $M_\text{mess}$ & $ m_{\tilde{g}}$ \quad & $m_{3/2}$ \, & $T_S $ & $T_\kappa$ & $T_\text{rh}$ \,\,  &$\Omega_{3/2} h^2$
\\[0.5ex]
\hline \hline
A & $10^6$ &  $10^{12}$ & $10^{-3}$ &  $ > 10^{10}$ & $3 \times 10^3$ & $\geq 10^2$ &   $10^9 $ & $10^{11}$ &  $< 10^{10}$ & $\leq 0.12$
\\[0.5ex]
B & $10^4$ &  $10^{9}$ & $10^{-4}$ &  $> 10^{9}$ & $3 \times 10^3$ & $\geq 10$  &   $10^{8} $ & $10^{9}$ & $< 10^{11}$  & $\leq  0.12$
\\[0.5ex]
C & $10^3$ &  $10^{10}$ & $10^{-4}$ &  $> 10^{9}$ &  $3 \times 10^3$ &  $\geq 1$ &   $10^7 $ & $10^9$ & $< 10^{12}$ & $\leq 0.12$
\\[0.5ex]
D & $10^2$ &  $10^{7}$ & $10^{-5}$ &  $>10^{7}$ &  $5 \times 10^3$  &  $\geq 2.5$ &  $10^7 $ & $10^{7}$ & $< 10^{11.6}$  & $ \leq 0.12$
\\[0.5ex]
E & $10$ &  $10^{7}$ & $10^{-5}$ &  $> 10^{7}$ &  $3 \times 10^3$ &  $\geq 0.1$ &   $10^6 $ & $10^7$ & $< 10^{13}$ & $\leq 0.12$
\\[0.5ex]
F & $1$ &  $10^{5}$ & $10^{-6}$ &  $> 10^{6}$ &  $5 \times 10^3$ &  $\geq 0.25$ &   $10^6 $ & $10^5$ & $< 10^{12.6}$ & $\leq 0.12$
\\[0.5ex]
\hline \hline
\end{tabular}
\label{table}
\caption{List of benchmark values for the $m_S$, $S_0$, $\kappa_\varphi$, $M_\text{mess}$, $m_{\tilde{g}}$ and $m_{3/2}$ with the maximum reheating temperature value that give an acceptable gravitino abundance (the allowed combination of values in each line is not unique, see also figure 6b). The corresponding $T_S$ and $T_\kappa$ temperatures are also written. Numbers of order one 
are omitted. We considered the superpotential interaction (\ref{wtab}) with $n=1$; for $n>1$ similar results are obtained. All the dimensionful quantities are in GeV units.}  
\end{table}

\subsubsection{PQ-symmetry}
In the case of the Peccei-Quinn symmetry the charges are $PQ[S]=1$, $PQ[H_u]=-a/2$ and $PQ[H_d]=-a/2$. For $a>1$ the $S$ does not have any renormalizable interactions in the effective theory. An example is the Kim-Nilles mechanism \cite{Kim:1983dt} with $a=2$ and the VEV of $S$ between $10^9 \text{ GeV}\lesssim  S_0  \lesssim 10^{12}$ GeV.  The Peccei-Quinn symmetric GMSB solution to the $\mu/B\mu$ problem requires extra messenger fields through which the SUSY breaking is mediated to the PQ sector and radiatively stabilize the singlet $S$, see e.g. ref. \cite{Giudice:2007ca, Jeong:2011xu}. An example is given by the superpotential
\begin{equation} \label{pq2}
W=FX+\frac{S}{\Lambda_*}X \left(\phi\bar{\phi}+\phi\bar{\psi} \right)+ \kappa_\psi S \left(\psi\bar{\psi} + \psi\bar{\phi}\right) + \kappa_H\frac{S^2}{M_\text{Pl}}H_uH_d +M_{\text{mess}(\phi)}\phi\bar{\phi}+M_{\text{mess}(\psi)}\psi\bar{\psi}
 \end{equation}
where $\phi+ \bar{\phi}$ and $\psi+ \bar{\psi}$ account for the two sectors of messenger fields that belong to $5+\bar{5}$ representations of the SU(5) into which the SM gauge groups are embedded.
The mass $m_S$ of the scalar component of $S$ (called saxion)  is generally of the order of the MSSM soft masses, while the fermionic component (axino) is much lighter. Under certain conditions the axino can be the LSP. Nonetheless, still the abundance of the unstable gravitino has to be much constrained because it will decay to hot dark matter axinos.

In the model (\ref{pq2}),  the scalar $S$ receives a plasmon mass from the Higgses and the messengers at high enough temperatures and gets thermally displaced to the origin. Due to $M_\text{Pl}$ suppressed non-renormalizable couplings to Higgs bilinear the thermal 
effects coming from the interaction $\delta W= (S^2/M_\text{Pl})H_u H_d$ are negligible below the temperature $({m_S M_\text{Pl}})^{1/2}$. 
If the mass of the messengers $\psi$, $\bar{\psi}$ is less than $({m_S M_\text{Pl}})^{1/2}$  the leading thermal effects on the singlet potential are due to the messengers $\psi$, $\bar{\psi}$. 
The critical temperature $T_S$ can be as low as the freeze-out temperature of the messengers $M_{\text{mess} (\psi)}/20$. 
It is actually 
\begin{equation}
T_S=\text{max}\left\{\frac{m_S}{\kappa_\psi}, \, M_{\text{mess}(\psi)}\right\}
\end{equation}
and the Goldstino production from thermal scattering of messengers fields $\phi$, $\bar{\phi}$ changes as described in the expression (\ref{Oepsilon}) with the suppression parameter $\epsilon$ determined by the ratio $T_S/M_{\text{mess}(\phi)}$. Apparently, the mechanism works for $M_{\text{mess}(\psi)}<M_{\text{mess}(\phi)}$ and coupling $\kappa_\psi$ sufficiently small (\ref{1}) so that $T_\kappa<M_{\text{mess}(\phi)}$. The last requirement, $\kappa_\psi \ll 1$, also prevents thermal inflation or an $S$-domination phase from taking place. In the case that the thermal dispersion of the singlet is significant and triggers the transition to $S_0$, then one should ask for $T_\kappa \sim S_0$ to avoid overproduction -see also the discussion for the NMSSM case below.  Typically, the thermal production of gravitinos effectively initiates at $T_S$ 
and  the relic abundance of LSP gravitinos is given by the expression (\ref{OgA}) even though the reheating temperature might be well above the messenger scales, see Table 2.

\subsubsection{The NMSSM}

Instead of the PQ-symmetry, the discrete $Z_3$ symmetry can forbid the dimensionful $\mu$-term which is dynamically generated by the NMSSM singlet, see \cite{Ellwanger:2009dp} for a review. 
In our case, an example of this scenario is described by the superpotential 
\begin{equation} \label{Wh}
W= FX+ \left(\frac{S}{\Lambda_*}\right)^n X\phi\bar\phi + \kappa_H S H_u H_d+\frac{\kappa'}{3} S^3 +M_\text{mess} \phi\bar{\phi} \,,
\end{equation}
where the power $n$ is introduced in order invariance under symmetries to be satisfied by the non-renormalizable term. The $Z_3$  guarantees the form of the superpotential and the absence of a bare $\mu$-term, e.g. with $Z_3[S]=Z_3[\phi]=Z_3[H_u]=Z_3[H_d]=1/3$, $Z_3[\bar{\phi}]=-1/3$, $Z_3[X]=0$ and $n=3$.  The $Z_3$ is spontaneously broken and unless additional  $Z_3$ violating terms are present cosmologically problematic domain walls are formed \cite{Abel:1995wk}.  
Terms of the form $m^3_\text{soft}S$, proposed to heal the cosmological domain wall problem \cite{Panagiotakopoulos:1998yw} can be also considered, see ref. \cite{Kadota:2015dza} for a recent study.

The $\mu$-term is given by the relation $\mu=\kappa_H S_0$. The renormalizable interaction between the singlet and the Higgs bilinear thermalize the singlet at high temperatures, thus amplifying the gravitino production rate unless the coupling $\kappa_H$ is sufficiently small, as described in the condition (\ref{1}). This constraint implies that the VEV of the singlet has to be rather large, $S_0=\mu/\kappa_H$. 

When the singlet $S$ is identified with the NMSSM singlet the thermal dispersion $\delta S_T\simeq T_\kappa/\sqrt{24}$ plays a key r\^ole. 
This is because if $\left\langle S \right\rangle =0$ the gauge mediation contribution to soft terms vanishes. 
Vanishing soft terms prevent the transition to the zero temperature VEV $S_0$ from taking place\footnote{This might happen in the PQ-symmetry scenarios too.}. However, the thermal dispersion of the singlet generates an effective messenger coupling that is sufficient to trigger the transition to $S_0$ at the temperature $T_S\sim T_\kappa$,  where
\begin{equation}
T_S=\frac{|m_S|}{\kappa_H}\,
\end{equation}
if $T_\kappa \gtrsim S_0$   \footnote{ If $T_\kappa <S_0$, the supergravity effects and/or $Z_3$ violating supersymmetric terms could trigger the transition to $S_0$ but we will not elaborate on this case.}. This last requirement implies that the thermal dispersion contribution to the gravitino yield is not negligible. For $n=3$, $\lambda_\text{mess}=(S/\Lambda_*)^3$,  the gravitino abundance reads (\ref{OTMSSM})
\begin{equation} 
\left. \Omega^{\text{MSSM(sc)}}_{3/2}h^2\right|_{\delta S_T}\, \sim \, \frac{0.1}{7\,(24)^3} \left( \frac{T_\kappa}{S_0} \right)^{6}   \left( \frac{T_\kappa}{10^8 \,\text{GeV}}\right) \left(\frac{\text{GeV}}{m_{3/2}} \right) \left(\frac{m_{\tilde{g}}}{\text{TeV}}\right)^2 \,.
\end{equation}
From the above expression it is clear that the cosmologically most attractive case is to have $T_\kappa$ as small as possible in order to avoid gravitino overabundance, which together with the requirement $T_\kappa \gtrsim S_0$ results in  $T_\kappa \sim S_0$.
In order the messenger fields not to amplify the gravitino yield we require, as always, $T_\kappa <M_\text{mess}$. 
For  $T_\kappa \sim S_0$ and $T_S$ not much smaller than $T_\kappa $ the total gravitino yield receives also a sizable  contribution from  the non-thermal part, that is 
\begin{equation} 
\left. \left.  \Omega^{\text{MSSM(sc)}}_{3/2}h^2 \sim \Omega^{\text{MSSM(sc)}}_{3/2}h^2\right|_{\delta S_T} + \Omega^{\text{MSSM(sc)}}_{3/2}h^2\right|_{\left\langle S(T) \right\rangle} \,.
\end{equation}
Gravitinos are efficiently produced thermally below the temperature $T_\kappa \sim T_S=|m_S|/\kappa_H$.
At higher temperatures neither the MSSM sector nor the secluded GMSB sector generate helicity $\pm1/2$ gravitino modes because 
it is $\left\langle \lambda_\text{mess}(T)\right\rangle \sim 0$ and the integral (\ref{YT}) vanishes. Thereby, viable GMSB cosmology with high reheating temperatures is possible, see Table 3.

\begin{table}
\centering 
\begin{tabular}{||r|r|r|r|r|r|r||r|r||r||r||} 
\hline \hline
 & {$|m_S|$} &  $S_0$ & $\kappa_H\,$ & $M_\text{mess}$ & $ m_{\tilde{g}}$ \quad & $m_{3/2}$ \, & $T_S $ & $T_\kappa$ & $T_\text{rh}$ \,\,  &$\Omega_{3/2} h^2$
\\[0.5ex]
\hline \hline
G & $10^3$ &  $10^{7}$ & $10^{-5}$ &  $>10^{8}$ &  $3 \times 10^3$  &  $\geq 1$ &  $ \lesssim T_\kappa $ & $10^{7}$ & $< 10^{12}$  & $ \leq 0.12$
\\[0.5ex]
H & $10^3$ &  $10^{5}$ & $10^{-6}$ &  $> 10^{5}$ &  $3 \times 10^3$ &  $\geq 0.01$ &   $\lesssim T_\kappa $ & $10^5$ & $< 10^{14}$ & $\leq 0.12$
\\[0.5ex]
I & $10^2$ &  $10^{5}$ & $10^{-6}$ &  $> 10^{5}$ &  $3 \times 10^3$ &  $\geq 0.01$ &   $\lesssim T_\kappa $ & $10^5$ & $< 10^{14}$ & $\leq 0.12$
\\[0.5ex]
J & $10$ &  $10^{5}$ & $10^{-6}$ &  $> 10^{5}$ &  $3 \times 10^3$ &  $\geq 0.01$ &   $\lesssim T_\kappa $ & $10^5$ & $< 10^{14}$ & $\leq 0.12$
\\[0.5ex]
\hline \hline
\end{tabular}
\label{table2}
\caption{List of benchmark values, as in Table 2, for superpotential interaction (\ref{Wh}) with $n=3$. Here, the transition to $S_0$ is
triggered at the temperature $T_\kappa$ and it is $m^2_S<0$ .
All the dimensionful quantities are in GeV units.}  
\end{table}

The most interesting implication of choosing the NMSSM singlet to generate dynamically the spurion-messenger coupling  is that  the NMSSM singlet sector is subject to existing phenomenological constraints and
it is a promising window to test the mechanism discussed in this paper. 

The first constraint originates from the fact that $\mu$-term is given by the relation $\mu=\kappa_H S_0$ and has to be above about 100 GeV in order
to be in agreement with null searches for charginos at LEP \cite{Abdallah:2003xe}. This implies a lower bound on $\kappa_H$ for a given value of $S_0$. 
Moreover, the magnitude of the mixing of the singlet with the 125 GeV Higgs, which is proportional to $\kappa_H^2 S_0v/(m_S^2-m_h^2)$,  is
constrained from above by the data \cite{Badziak:2013bda}. On one hand, this mixing has to be small enough to be consistent with the LHC Higgs
coupling measurements. On the other hand, if the singlet is light enough it would have already been seen either at LEP or LHC if the Higgs-singlet
mixing is too large. Therefore, for a given value of the singlet mass there is an upper bound on  $\kappa_H^2 S_0$. In spite of these constraints, it
is well possible to have cosmologically viable gravitino dark matter and satisfy all the experimental constraints even if the singlet is light, as
demonstrated in Table 3.

In workable scenarios of NMSSM GMSB one has to go beyond the minimal models described by the superpotential (\ref{Wh}). 
Usually, the minimal model is extended to include appropriate couplings between the NMSSM singlet $S$ and other fields 
\cite{Dine:1993yw, Delgado:2007rz, Hamaguchi:2011kt, Allanach:2015cia}
via renormalizable interactions.   
This way the singlet $S$ gets a sizable negative mass squared value resulting in a sufficiently large $S_0$ VEV and a successful electroweak symmetry breaking.
In some scenarios it is $m_S \sim 100$ GeV, see e.g. \cite{Allanach:2015cia}, hence the $T_S$ is much larger than the electroweak scale, see Table 3.

\section{Conclusions}

If the gravitino is the stable LSP then it is a component of the dark matter in the universe and the estimation of its relic density is of central cosmological importance. 
Both the MSSM and the messenger fields of the GMSB secluded sector contribute to the gravitino production and their properties should be known in detail.  
Except for a particular finely tuned combination of masses and reheating temperatures the gravitino is generally found to be overabundant putting stringent constraints on many GMSB models. If TeV scale SUSY is realized in nature, the constrained gravitino cosmology can actually provide us with an insight into the secluded/hidden sector building blocks.

In this paper we elaborated on the idea that the messenger coupling $\lambda_\text{mess}$ is controlled by the VEV of a new field,  $S$, according to the relation  $\lambda_\text{mess}(S)=(S/\Lambda_*)^n$ where  $\Lambda_*$ a cut-off scale.
This legitimate assumption is of special interest both for its model building and cosmological aspects and in this work we  focused on the cosmological implications. 
The underlying motivation for such a study is the fact that the messenger Yukawa coupling $\lambda_\text{mess}$  is a key quantity for the determination of the relic gravitino yield hence, any dynamical change of its value modifies the production rate of the gravitino. When the VEV of $S$ goes to zero the messengers decouple from the SUSY breaking spurion $X$ and the production of the helicity $\pm1/2$ gravitino modes gets effectively switched off. 
 Apparently, a vanishing messenger coupling is unwanted at zero temperature because it cancels the GMSB effects, however at finite temperature the $\lambda_\text{mess}(T) \rightarrow 0$ assumption is a reasonable and phenomenologically valid one.

In the scenario where $\lambda_\text{mess}=(S/\Lambda_*)^n$ the gravitino cosmology is possible to be much less sensitive to the reheating temperature thanks to the dynamic nature of the messenger coupling value. Nonetheless, adopting a $\lambda_\text{mess}(S)$ messenger coupling does not automatically imply a suppressed gravitino production rate. The singlet is also required to be sufficiently weakly coupled to the thermal plasma in order not to get thermalized too fast and, in some cases, has to be stabilized in an intermediate energy scale. Otherwise, the $S$-dependent messenger coupling may lead to opposite conclusions, regarding the gravitino cosmology, since the thermal dispersion of the singlet can amplify the gravitino production rate. In addition, if it is $\kappa\sim 1$ the thermal trapping of the singlet may cause a late entropy production which may wash out effects taking place at higher temperatures. The requirement $\kappa \ll 1$ might look, at first sight, an unattractive feature of the mechanism  but 
in principle it can be explained and embedded in technically natural models. The gravitino mass window that is favored in the $\lambda_\text{mess}(T)$ scenario spans from about $10^{-2}$ GeV to 100 GeV.

An intriguing possibility is to couple the singlet $S$ to the Higgs bilinear in the superpotential and charge it under a non-trivial symmetry such as the Peccei-Quinn or the $Z_3$ ones. In the first case the coupling to Higgs bilinear is non-renormalizable whereas in the second is renormalizable with the $S$ superfield identified with the NMSSM singlet. In both cases the singlet, for phenomenological reasons, is originally considered to have additional couplings to vector-like pairs of messenger fields.  These scenarios are indeed interesting possibilities. 

When the messenger coupling vanishes at high temperatures the relic gravitino abundnace is determined by the characteristic temperature  $T_S$ of the $S$ transition to the zero temperature minimum $S_0$ since the gravitino yield is rendered  insensitive to the thermalization of the messenger fields. We mention that thermalized messengers can also explain the selection of the SUSY breaking vacuum \cite{Dalianis:2010pq}. 
The result that $\Omega_{3/2}h^2\propto T_S$ significantly relaxes the gravitino cosmology which is much constrained due to the increased lower bounds on the gluino masses from the LHC searches. For example, an updated survey of the allowed General Gauge Mediation parameter space \cite{Knapen:2015qba} indicates gluinos with masses $3-6$ GeV. In our scenario, neither the heavy gluino masses nor the messenger mass scale impose a direct bound on the reheating temperature. 
Thereby, high enough reheating temperatures for leptogenesis to take place and viable gravitino cosmology can be reconciled when a field-dependent messenger Yukawa coupling is considered. 

The mechanism presented here is a simple generalization of the minimal GMSB models and underlines the strong link between the $\Omega_{3/2}$ and the secluded/hidden sector dynamics with substantial cosmological and potentially rich phenomenological implications.

\section*{Acknowledgments}
This work was partially supported by Polish National Science Centre 
under research grants DEC-2012/05/B/ST2/02597, DEC-2014/15/B/ST2/02157
and DEC-2012/04/A/ST2/00099. MB acknowledges support from the Polish 
Ministry of Science and Higher Education (decision no.\ 1266/MOB/IV/2015/0).
MB thanks the Galileo Galilei Institute for Theoretical Physics and INFN for hospitality and partial support during the completion of this work.

\appendix

\section{Elements of gravitino production from thermal scatterings}
Spontaneous breaking of supersymmetry gives rise to a massles spin-$1/2$ Goldstino, the Nambu-Goldstone mode associated with the supersymmetry breaking. In the context of supergravity the Goldstino is absorbed as the longitudinal component of the spin-$3/2$ gravitino, giving rise to a spin-$3/2$ particle with four helicity states and mass $m_{3/2} =F/(\sqrt{3} M_\text{Pl})$. We have considered that the value of the $X$ auxiliary field is mainly responsible for the size of the supersymmetry breaking scale, $F\simeq F_X$. In the low-scale supergravity models such as the gauge mediated supersymmetry breaking models the gravitino is naturally the LSP. 

The two helicities, $\pm1/2$ and $\pm3/2$ have a different abundance. The abundance of the gravitationally interacting helicity $\pm3/2$ gravitinos depends universally on the maximum reheating temperature \cite{Ellis:1984eq, Ellis:1984er}
\begin{equation}
\Omega^\text{grav}_{3/2}h^2 \sim 0.1 \left( \frac{m_{3/2}}{\text{GeV}}\right) \left(\frac{T_\text{rh}}{10^{12}\text{GeV}} \right) \,.
\end{equation}

\subsection{MSSM contribution}

The MSSM predicts a temperature dependent gravitino abundance (\ref{MSSM}) unless the gravitino is thermalized. The gravitno thermal abundance is $\Omega^{\text{th}}_{3/2} h^2 \sim (m_{3/2}/\text{keV})$. If sparticles are abundant in the plasma the $\pm1/2$ helicity gravitinos can reach thermal equilibrium above the temperature
\begin{equation}
T^f_{3/2} \sim \,2 \times 10^{5}\, \text{TeV} \left( \frac{m_{3/2}}{\text{MeV}} \right)^2 \left(\frac{\text{TeV}}{m_{\tilde{g}}} \right)\,.
\end{equation}
A very light gravitino can also reach thermal equilibrium with Standard Model light degrees of freedom, e.g. via gravitino pair annihilation into two photons, even for temperatures below the sparticle masses \cite{Drees:2004jm}. 
It is found that $T^f_{3/2} <{\cal O}$(TeV) for gravitinos with mass $m_{3/2}< {\cal O}(1)$eV. Gravitinos do not intervene  with the Big Bang Nucleosynthesis if they decouple before muons did, that is $T^f_{3/2} \gtrsim 100$ MeV, which implies $m_{3/2} \gtrsim 10^{-4}$eV. 
Taking also into account the large scale structure data \cite{Viel:2005qj}, a gravitino with mass $m_{3/2}<1$ keV accounts for Warm Dark Matter (WDM) component. The observed matter power spectrum constrains the WDM contribution to be less than $12\%$ to the total dark matter abundance which translates into an upper bound for the WDM gravitino to be $m_{3/2}<0.16$ eV.
Thereby, one finds that gravitinos with mass $10^{-4} \text{eV}<m_{3/2}<16$ eV are cosmologically safe. 
This cosmologically accepted light-gravitino mass window is derived considering freezing out of thermalized gravitinos, see figure 2.

We remind the reader that the formuli for the relic abundance of helicity $\pm1/2$ gravitinos written in the paper are valid only for gravitino freezing out temperature larger than $M_\text{mess}$ or $T_\text{mess}$, that is not for too light gravitinos.


\subsection{Messenger contribution}
The cross section of the scattering processs that generate Goldstinos and involve messenger particles $j,j'$ reads \cite{Choi:1999xm}, 
\begin{equation} \label{app1}
\begin{split}
\sum_{i, \, j,\, j'}  
&\sigma(i+j   \rightarrow j' + \psi_G)\, (\bar{s})  =\\
& \xi\, \lambda_\text{mess}^2(\bar{s}) \, \frac{2(2\bar{s}^2-3\bar{s}M^2_\text{mess}+M^4_\text{mess})+\bar{s}(\bar{s}-2M^2_\text{mess})\log(\frac{\bar{s}^2}{M^4_\text{mess}}) }{\bar{s}(\bar{s}-M^2_\text{mess})^2} \\
& \equiv \xi\, \lambda_\text{mess}^2(\bar{s}) \, f_{ij}(\bar{s}, M_\text{mess})
\end{split}
\end{equation}
and
\begin{equation} \label{cs2}
\begin{split}
\frac{1}{2} \sum_{j,\, j', \, i} 
&\sigma(j+j' \rightarrow i + \psi_G)\,(\bar{s}) =\\
& \xi \lambda_\text{mess}^2(\bar{s})\, \frac{2[\bar{s}(\bar{s}-4M^2_\text{mess})]^{1/2}+M^2_\text{mess}\log\left(\frac{\bar{s}-2M^2_\text{mess}-[\bar{s}(\bar{s}-4M^2_\text{mess})]^{1/2}}{\bar{s}-2M^2_\text{mess}+[\bar{s}(\bar{s}-4M^2_\text{mess})]^{1/2}} \right)}{\bar{s}(\bar{s}-4M^2_\text{mess})}\\
& \equiv \xi\, \lambda_\text{mess}^2(\bar{s}) \, f_{jj'}(\bar{s}, M_\text{mess})\,,
\end{split}
\end{equation}
where $i$ corresponds to a MSSM particle and $\bar{s}$ is the center of mass energy squared.
The scattering yield of gravitinos, i.e. the number to entropy ratio, is given by the expression
\begin{align}
Y^\text{mess(sc)}_{3/2}
\begin{split}
& = \int^{T_\text{rh}}_{T} \frac{\left\langle \sigma_{1+2\rightarrow 3+ \psi_G} \, v_{12} \, n_1\,  n_2 \right\rangle}{\varsigma(T)H(T)T} dT \\
& =\frac{\bar{g}M_\text{Pl}}{16\pi^4} \int^\infty_{x_\text{rh}}dx \, x^3 K_1(x) \int^{xT_\text{rh}}_{M_i+M_j} d\sqrt{\bar{s}}\,\sigma(\bar{s}) \left[\frac{(\bar{s}-M^2_1-M^2_2)^2-4M^2_1M^2_2}{\bar{s}^2} \right] 
\end{split}
\end{align}
with $x=(M_1+M_2)/T$. 
Considering that the incident particles can be either messengers, $M_{j,j'}=M_\text{mess}$, or MSSM particles which are nearly massles, $M_i \simeq 0$,  we take the yield
\begin{equation}
\begin{split}
Y^\text{mess(sc)}_{3/2}\, =\, \frac{\bar{g}M_\text{Pl} \xi}{16\pi^4} 
& \int^\infty_{x_\text{rh}}dx \, x^3  K_1(x) \times  \left[ \int^{xT_\text{rh}}_{M_\text{mess}} d\sqrt{\bar{s}}\, \left\langle \lambda_\text{mess}^2(\bar{s}) \right\rangle \,f_{ij}(\bar{s}) \frac{\left(\bar{s}-M^2_\text{mess} \right)^2}{\bar{s}^2}\, +\right. \\ 
&\left. +\, \int^{xT_\text{rh}}_{2M_\text{mess}} d\sqrt{\bar{s}}\, \left\langle \lambda_\text{mess}^2(\bar{s}) \right\rangle \,f_{jj'}(\bar{\bar{s}}) \frac{\left(\bar{s}-4M^2_\text{mess} \right)^2}{\bar{s}^2} \right]\,.
\end{split}
\end{equation}
We find that the contribution to the yield from the first integral in the brackets, corresponding to the case where one of the incident particles is a messenger, is about $8\pi$ larger than the second one when $T_\text{rh} \gg M_\text{mess}$. 

For constant messenger coupling $\lambda_{\text{mess}\,0}$ we find 
\begin{equation} \label{Yscat}
\begin{split} Y^\text{mess(sc)}_{3/2} \, & 
\left. \simeq   \frac{124 \pi}{15} \frac{T^6}{16\pi^4\,\varsigma(T)H(T)T} \right|_{T=M_\text{mess}} \xi \, \lambda_{\text{mess}\,0}^2  \frac{1}{M_\text{mess}} \\
& \simeq \, 8 \xi \times 10^{-5} \lambda_{\text{mess}\,0}^2  \left[\frac{270}{g_*(T_\text{rh})} \right]^{3/2}\frac{M_\text{Pl}}{M_\text{mess}}\,,
\end{split}
\end{equation}
and the relic gravitino abundance from messenger scatterings is found to be
\begin{equation} 
\Omega^\text{mess(sc)}_{3/2} \, h^2 
 \sim \, 0.6\,\xi\, \left(\frac{10^5\,\text{GeV}}{M_\text{mess}}\right)\left(\frac{m_{3/2}}{\text{GeV}}\right) \left(\frac{\lambda_{\text{mess}\,0}}{10^{-9}} \right)^2 \left(\frac{270}{g_*(T_\text{rh})}\right)^{3/2}\,.
\end{equation}
Writing the messenger coupling in terms of the gaugino mass we attain the expression (\ref{mess1}) given in the introduction.

\end{document}